\documentclass[a4paper,11pt]{article}
\usepackage{jheppub} 
\usepackage{lineno}
\usepackage{subfig}
\usepackage{booktabs}
\usepackage{makecell}

\newcommand{\omits}[1]{}

\def\bc{\begin{center}}

\def\ec{\end{center}}
\def\be{\begin{eqnarray}}
\def\ee{\end{eqnarray}}

\title{\boldmath Island of Reissner-Nordstr{\"o}m anti-de Sitter black holes in the large $d$ limit}

\author[a]{Chen-Wei Tong,}
\author[a]{ Dong-Hui Du,}
\author[a]{ Jia-Rui Sun}
\affiliation[a]{School of Physics and Astronomy, Sun Yat-Sen University, Guangzhou 510275, China}
\emailAdd{tongchw3@mail2.sysu.edu.cn, donghuiduchn@gmail.com, sunjiarui@mail.sysu.edu.cn}

\abstract{We study the information paradox of Reissner-Nordstr{\"o}m anti-de Sitter (RN-AdS$_{d+1}$) black holes in the large dimension limit by using the island formula. The entanglement entropy of Hawking radiation is calculated both for the non-extremal and the extremal cases, in which the boundary of the radiation region is close to the outer horizon. For the non-extremal case, the entanglement entropy of Hawking radiation obeys the Page curve, i.e. the entanglement entropy of Hawking radiation increases with time and reaches saturation about twice Bekenstein-Hawking entropy at the Page time. For the extremal case, the entanglement entropy of Hawking radiation becomes ill-defined in the absence of the island due to the appearance of the singularity at the origin of the radial coordinate, while when the island exists, the entanglement entropy is found to be equal to the Bekenstein-Hawking entropy. In addition, for the case where the boundary of the radiation region is close to the horizon,  there are some obvious constraints required by the existence of island solution for both non-extremal and extremal cases, which can be utilized to put constraints on the size of the black hole. These results reveal new features of the semi-classical large $d$ black holes from the island perspective.}

\begin{document}
	\maketitle
	\flushbottom
	
\section{Introduction}
\label{sec.1}

In 1974, Hawking discovered that the black hole can emit thermal radiation which is called Hawking radiation~\cite{Hawking:1974rv}. In this way, the black hole has a temperature and can evaporate. However, a consequent problem called the black hole information paradox~\cite{Hawking:1976ra} is noticed during the black hole evaporation in the semi-classical gravity: essentially, a black hole formed by the collapse of pure state becomes the mixed state after thermal Hawking radiation. Obviously, this process is not unitary evolution, which is in conflict with the standard rules of quantum mechanics. To avoid the information loss problem in the semi-classical gravity, Page proposed that the evolution of the entanglement entropy produced in the radiation process of the black hole should satisfy the Page curve \cite{Page:1993wv,Page:2013dx}. Therefore, the key point to solve the black hole information loss problem is to reproduce the Page curve in the semi-classical description.

Recently, significant progress has been made in the study of black hole information paradox in the light of the AdS/CFT correspondence~\cite{Maldacena:1997re}. In Refs. \cite{Penington:2019npb,Almheiri:2019psf,Almheiri:2019hni,Almheiri:2020cfm}, the island formula was proposed to calculate the entanglement entropy of Hawking radiation, which helped to reproduce the Page curve in semi-classical gravity. This formula comes from the holographic entanglement entropy(HEE) formula (or RT formula)~\cite{Ryu:2006bv,Hubeny:2007xt} and its quantum version, i.e. the quantum extremal surface (QES)~\cite{Faulkner:2013ana,Engelhardt:2014gca}. It is worth noting that the island formula can be derived through the gravitational path integral formalism \cite{Penington:2019kki,Almheiri:2019qdq}. The island formula for the entanglement entropy of Hawking radiation is given by
\begin{equation}
	\label{eq1.2}
	S(R)=min\left \{ ext\left \{ S_{gen} \right \}  \right \} =min\left \{ ext\left \{ \frac{Area(\partial I)}{4G_{N}}+S_{\rm{matter}}^{\rm{finite}}(R\cup I)  \right \}  \right \}  ,
\end{equation}
where $I$ is the island and $R$ is the radiation region outside the black hole. Note that due to the short distance cut-off \cite{Bombelli:1986rw,Srednicki:1993im}, the entanglement entropy of matter has UV divergence, which can be absorbed by the renormalizing the Newton constant $G_{N}$ \cite{Susskind:1994sm}. $S_{gen}$ is the generalized entropy composed of two parts: the first area term is the contribution of the island, and the second term is the finite part of the entanglement entropy of the matter on the union of the radiation region and the island. The entanglement island or simply called island is a region in the black hole interior, which is found to be a part
of the entanglement wedge of the Hawking radiation outside the black hole. The island formula is expected to applicable to different types of black holes. So far, the island formula has been applied into the (1+1)-dimensional gravitational models \cite{Almheiri:2019yqk,Chen:2019uhq,Chen:2019iro,Marolf:2020xie,Chen:2020wiq,Anderson:2020vwi,Gautason:2020tmk,Hartman:2020swn,Anegawa:2020ezn,Hollowood:2020cou,Goto:2020wnk,RoyChowdhury:2022awr,Djordjevic:2022qdk,Yu:2022xlh,Wu:2023uyb,Hashimoto:2020cas}, and also some higher dimensional models \cite{Almheiri:2019psy,Hashimoto:2020cas,Alishahiha:2020qza,Ling:2020laa,Matsuo:2020ypv,Wang:2021woy,Karananas:2020fwx,Kim:2021gzd,Yu:2021cgi,Ahn:2021chg,He:2021mst,Arefeva:2021kfx,Yu:2021rfg,Gan:2022jay,Yadav:2022fmo,Azarnia:2022kmp,Anand:2022mla,Du:2022vvg,HosseiniMansoori:2022hok,Guo:2023gfa}. Besides, there were also studies on equivalent description of the entanglement islands based on the AdS/BCFT correspondence \cite{Takayanagi:2011zk,Fujita:2011fp}, see \cite{Geng:2020qvw,Chen:2020hmv,Geng:2020fxl,Hu:2022ymx,Suzuki:2022xwv,Anous:2022wqh,Lin:2022aqf,Miao:2022mdx,Miao:2023unv,Li:2023fly} for examples.

In the present paper, we mainly consider the entanglement island of RN-AdS$_{d+1}$ black hole coupled an auxiliary thermal bath in the large $d$ limit for both non-extremal and extremal cases, in which the gravitational theory shows interesting decoupling properties. For a large $d$ black hole, its near horizon geometry will be driven into a fixed point and its dynamics will decouple with that in the far region, and the effective theory reduces to a 2-dimensional dilaton gravity. In addition, the gravitational perturbations will go into the hydrodynamic limit, which makes analytical calculations for the Equations of motion (EoMs) of the gravitational perturbations available~\cite{Emparan:2013moa,Chen:2015fuf,Guo:2015swu,Chen:2017wpf,Chen:2018vbv,Emparan:2020inr,Suzuki:2022apk,Luna:2022tgh}. Thus large $d$ black holes provide good platform for studying the entanglement island analytically. On the other hand, the properties of entanglement island for extremal black holes have not been completely understood yet. Previously works attempted to use the island formula to analyze the extremal black hole, but their calculations were mainly based on taking extremal limit from the non-extremal case~\cite{Yu:2021cgi,Karananas:2020fwx}. Subsequent studies suggested that  the non-extremal and extremal cases should be analyzed separately based on the their different Penrose diagrams~\cite{Kim:2021gzd,Ahn:2021chg,Carroll:2009maa}. Moreover, although the Hawking temperature of a extremal black hole is zero, its Bekenstein-Hawking area entropy is non-zero, which also has been verified from counting microstates of solitons for extremal black holes in string theory \cite{Strominger:1996sh} and calculating the microscopic entropy of the CFT holographically dual to the extremal black hole~\cite{Strominger:1997eq}. Now for the large $d$ the RN-AdS$_{d+1}$ black hole, it has been shown that it contains new dual CFT description in the (near) extremal limit. Therefore, it is interesting to further study the entanglement island of the large $d$ RN-AdS$_{d+1}$ black hole, which will give deeper understanding on the microscopic entropy and entanglement property of the extremal black holes.

This paper is organized as follows. In Sec. \ref{sec.2}, we will briefly review RN-Ad$S_{d+1}$ black holes in the large $d$ limit, both for the non-extremal and extremal cases. In Sec. \ref{sec.3}, we review the formulas to calculate the entanglement entropy of matter. In Sec. \ref{sec.4} and Sec. \ref{sec.5}, we mainly study the entanglement entropy of Hawking radiation for the non-extremal and extremal cases by using the island formula. In Sec. \ref{sec.6}, we discuss the constraints in the presence of island in more detail. In Sec. \ref{sec.7}, the Page curve and Page time are discussed. Finally, Conclusion and discussion are given in Sec. \ref{sec.8}.
\section{Review of the RN-AdS$_{d+1}$ black hole }
	\label{sec.2}
	In this section, we review the RN-AdS$_{d+1}$ black hole and its metric in large $d$ limit~\cite{Guo:2015swu}. Based on the previous results, we rewrite the metric for the non-extremal and extremal cases in the Kruskal coordinates.
	
The action of ($d+1$)-dimensional Einstein-Maxwell theory has the form as
	\begin{equation}
		I=\frac{1}{16 \pi G_{N}} \int d^{d+1} x \sqrt{-g}\left(R+\frac{d(d-1)}{L^{2}}-\frac{L^{2}}{g_{s}^{2}} F_{\mu \nu} F^{\mu \nu}\right),
	\end{equation}
	where $G_{N}$ is the Newton constant, $R$ is the Ricci scalar, $L$ is the curvature radius of the asymptotically AdS$_{d+1}$ spacetime and $g_{s}$ is the dimensionless coupling constant of U(1) gauge field. The equations of motion can be found as
	\begin{equation}
		\begin{aligned}
			R_{\mu \nu}-\frac{1}{2} g_{\mu \nu} R-\frac{d(d-1)}{2 L^{2}} g_{\mu \nu} & =\frac{L^{2}}{2 g_{s}^{2}}\left(4 F_{\mu \lambda} F_{\nu}{ }^{\lambda}-g_{\mu \nu} F_{\alpha \beta} F^{\alpha \beta}\right) \\
			\partial_{\mu}\left(\sqrt{-g} F^{\mu \nu}\right) & =0 ,
		\end{aligned}
	\end{equation}
	which admits the following RN-AdS$_{d+1}$ black hole solution
	\begin{equation}
		\label{eq2.3}
		\begin{aligned}
			d s^{2}&=-f(r) d t^{2}+\frac{d r^{2}}{f(r)}+r^{2} d \Omega_{d-1}^{2},\\
			A&=\mu \left(1-\frac{r_{+}^{d-2}}{r^{d-2}} \right)dt,
		\end{aligned}
	\end{equation}
	with
	\begin{equation}
		\begin{aligned}
			f(r)&=1-\frac{M}{r^{d-2}} +\frac{Q^{2}}{r^{2d-4}} +\frac{r^{2}}{L^{2}}, \\
			\mu&= \sqrt{\frac{d-1}{2(d-2)}\frac{g_{s}Q}{Lr_{+}^{d-2}}  }.
		\end{aligned}
	\end{equation}
	where $r_{+}$ is the radius of the outer horizon, $\mu$ is the chemical potential, $M$ and $Q$ are mass and charge of the RN-AdS$_{d+1}$ black hole.

\subsection{ The non-extremal large $d$ RN-AdS$_{d+1}$ black hole }
The dual CFT description of the large $d$ RN-AdS$_{d+1}$ black hole has been studied in~\cite{Guo:2015swu}. To analyze the large $d$ RN-AdS$_{d+1}$ black hole, defining $\rho \equiv \frac{r^{d-2}}{M}$ and $M\equiv r_{o}^{d-2}$, and taking the large dimension limit $d\rightarrow\infty$ together with the near horizon limit $r-r_+\ll r_+$, then the metric (\ref{eq2.3}) becomes
	\begin{equation}\label{non-extremal RN}
		\begin{aligned}
			d s^{2} & = -f(\rho) d t^{2}+ \frac{r^{2} }{(d-2)^{2}\rho ^{2}f(\rho)}d\rho ^{2}  +r_{o} ^{2}\rho^{\frac{2}{d-2}} d \Omega_{d-1}^{2}\\
			A&=\mu \left(1-\frac{\rho _{+}}{\rho } \right)dt ,
		\end{aligned}
	\end{equation}
	where
	\begin{equation}
		\label{eq2.6}
		\begin{aligned}
			f(\rho) & = 1-\frac{1}{\rho}+\frac{Q^{2}}{M^{2}} \frac{1}{\rho^{2}}+\frac{r_{o}^{2}\rho^{\frac{2}{d-2} } }{L^{2}}\\
			&\simeq   1-\frac{1}{\rho}+\frac{Q^{2}}{M^{2}} \frac{1}{\rho^{2}}+\frac{r_{o}^{2}}{L^{2}}.
		\end{aligned}
	\end{equation}
	Note that in order to obtain eq. (\ref{non-extremal RN}) we have used $\rho ^{\frac{2}{d-2}}\simeq 1$ (for finite $\rho$) once for $d\rightarrow\infty$, but we still keep the exact form of $\rho ^{\frac{2}{d-2}}$ that appears in the spherical term of the metric\footnote{Note that in the sphere radius $r_{o}\rho^{\frac{1}{d-2}}$ we are retaining an apparent term $\rho^{\frac{1}{d-2}}$ since it provides a contribution whenever the area element $r_{o}^{d-1}\rho^{\frac{d-1}{d-2}} \simeq r_{o}^{d-1}\rho$ is involved~\cite{Emparan:2020inr}. Essentially, this detail is crucial when we use the eq.~(\ref{eq3.2}) calculate the entanglement entropy. As you can see from the eq.~(\ref{eq4.6}), if we omit this term $\rho^{\frac{1}{d-2}}$, the location of the island can not be determined.}. The inner and outer horizon radius of large $d$ RN-AdS$_{d+1}$ black hole are
	\begin{equation}
		\rho _{\pm }=\frac{1\pm \sqrt{1-\frac{4kQ^{2}}{M^{2}} } }{2k},
		\quad  k=1+\frac{r_{o}^{2}}{L^{2}} .
	\end{equation}
	Thus the metric (\ref{non-extremal RN}) can be rewritten as
	\begin{equation}\label{ldnrn}
		ds^{2}=-\frac{k(\rho -\rho _{+})(\rho -\rho _{-})}{\rho ^{2}} d t^{2} +\left(k(\rho -\rho _{+})(\rho -\rho _{-}) \left(\frac{d-2}{r_{o}}\right)^{2}\right)^{-1}d\rho^2 +r_{o}^{2}\rho^{\frac{2}{d-2}} d \Omega_{d-1}^{2} .
	\end{equation}
	The temperature and entropy associating with the outer horizon are respectively
	\begin{equation}
		T=\left(\frac{d-2}{r_{o}}\right)\frac{k(\rho _{+}-\rho _{-})}{4\pi \rho _{+}}, \quad S_{BH}=\frac{\Omega _{d-1}r_{o}^{d-1}\rho _{+}}{4G_{N}},
	\end{equation}
	where $\Omega _{d-1}=2\pi^{d/2 }/\Gamma(d/2)$ is the volume of the unit sphere $S^{d-1}$. Besides, the tortoise coordinate is
	\begin{equation}
		\begin{aligned}
			\rho _{\ast }&=\int ^{\rho }\frac{r_{o}\rho d\rho}{(d-2)k(\rho -\rho _{+})(\rho -\rho _{-})}\\
			& =\frac{1}{2\kappa _{+}}\log(\rho -\rho _{+})+\frac{1}{2\kappa _{-}}\log(\rho -\rho _{-}),
		\end{aligned},
	\end{equation}
where $\kappa_{\pm}=\left(\frac{d-2}{r_o} \right)\left(\frac{k(\rho_+ -\rho_-)}{2\rho_{\pm}} \right)$ are the surface gravity on the inner and outer horizons, respectively. Then the Kruskal coordinates are
	\begin{equation}
		\label{eq2.14}
		U=-e^{-\kappa _{+}(t-\rho _{\ast } )} \quad , \quad  V=e^{\kappa _{+}(t+\rho _{\ast } )} .
	\end{equation}
	Finally we can rewrite metric (\ref{ldnrn}) in terms of the Kruskal coordinates as
	\begin{equation}
		\label{NonextKruskal}
		ds^{2}=-g^{2}(\rho)dUdV+r_{o}^{2} \rho^{\frac{2}{d-2}} d \Omega_{d-1}^{2} ,
	\end{equation}
	where
	\begin{equation}
		g^{2}(\rho)=\frac{k(\rho -\rho _{+})(\rho -\rho _{-})}{\rho ^{2}}\frac{1}{\kappa _{+}^{2}}e^{-2\kappa _{+}\rho _{\ast }} .
	\end{equation}

\subsection{ The extremal large $d$ RN-AdS$_{d+1}$ black hole }
\label{sec.2.2}
The extremal condition is $M^{2}=4kQ^{2}=r_{o}^{2d-4}$, in which the outer horizon coincide with the inner horizon  (i.e. $\rho_{+}=\rho_{-}=\frac{1}{2k}\equiv \rho_{h}$). Then the metric (\ref{ldnrn}) reduces to
	\begin{equation}\label{ldnrnex}
		ds^{2}=-\frac{k(\rho -\rho _{h})^{2}}{\rho ^{2}}  dt ^{2}+\left(k(\rho -\rho _{h})^{2} \left(\frac{d-2}{r_{o}}\right)^{2}\right)^{-1 }d\rho^2+r_{o}^{2}\rho^{\frac{2}{d-2}} d\Omega^2_{d-1},
	\end{equation}
and the tortoise coordinate now is
	\begin{equation}
		\begin{aligned}
			\rho_{\ast }&=\int ^{\rho }\frac{r_o\rho d\rho}{(d-2)k(\rho -\rho _{h})^2}\\
			&= \frac{r_{o}}{(d-2)k}\left [ -\frac{\rho _{h}}{\rho -\rho _{h}}+\log(\rho -\rho _{h})  \right ] ,
		\end{aligned}
	\end{equation}
	and the corresponding Kruskal coordinates are
	\begin{equation}
		\label{eq2.19}
		U=-e^{-\kappa (t-\rho_{\ast } )}, \quad V=e^{\kappa (t+\rho_{\ast } )} ,
	\end{equation}
	where we have still adopted the form of the Kruskal coordinates defined in eq. (\ref{eq2.14}). Note that in the extremal case, the surface gravity $\kappa_{+}=\kappa_-\equiv\kappa$ becomes zero, so we assume that the limit $\kappa \to 0$ to approach the final result of the extremal case. Now the metric (\ref{ldnrnex}) in terms of the Kruskal coordinates becomes
	\begin{equation}
		\label{extKruskal}
		ds^{2}=- w^{2}(\rho)dUdV+r_{o}^{2}d \Omega_{d-1}^{2} ,
	\end{equation}
	where
	\begin{equation}
		w^{2}(\rho )=\frac{k(\rho -\rho _{h})^{2}}{\rho ^{2}} \frac{1}{\kappa
			^{2}}e^{-2\kappa \rho _{\ast }} .
	\end{equation}
And the area entropy of the large $d$ extremal RN-AdS$_{d+1}$ black hole is
	\begin{equation}
		S_{BH}=\frac{\Omega_{d-1}r_{o}^{d-1}\rho_{h}}{4G_{N}}.
	\end{equation}

	\section{Formulas of the entanglement entropy of matter fields}
	\label{sec.3}
	In this section, in order to calculate the entanglement entropy of the radiation in any $(d+1)\ge4$ dimensional curved spacetime through the island formula~(\ref{eq1.2}), 	we will take some assumptions and adopt some limits to calculate the entanglement entropy of free massless matter field in higher-dimensional spacetime~\cite{Hashimoto:2020cas,Du:2022vvg}:
		\begin{itemize}
		\item If the distance between region $A$ and region $B$ is larger than the correlation length of the massive modes in the Kaluza-Klein tower of the spherical part, the Hawking radiation is assumed to be described by the two-dimensional s-wave modes (with the zero angular momentum) and influence from the higher angular momentum modes can be ignored. Then the finite part of the entanglement entropy of massless matter field is approximated by the mutual information of the two-dimensional massless fields as~\cite{Hashimoto:2020cas}
		\begin{equation}
			\label{EEFlat}
			S_{\rm{matter}}^{\rm{finite}}=-I(A:B)=\frac{c}{3}\log d(A,B) ,
		\end{equation}
		where $c$ is the central charge, $d(A,B)$ is the distance between the boundaries of region $A$ and region $B$ in flat spacetime. More specifically, in conformally flat spacetime the metric usually can be written in terms of the Kruskal coordinates :
		\begin{equation}
			ds^{2}=-\Omega^{2}dUdV,
		\end{equation}
		where $\Omega$ is conformal factor. Under a Weyl transformation, the distance $d(A,B)$ between two points in conformally flat spacetime can be written as~\cite{Gan:2022jay,Gautason:2020tmk,Hashimoto:2020cas}.
		\begin{equation}
			d(A,B)=\sqrt{\Omega(A)\Omega(B)[U(B)-U(A)][V(A)-V(B)]}.
		\end{equation}
		Therefore, the finite part of the entanglement entropy of matter fields is given by
		\begin{equation}
			\label{eq3.1}
			S_{\rm{matter}}^{\rm{finite}}=\frac{c}{6}\log\Omega(A)\Omega(B)[U(B)-U(A)][V(A)-V(B)].
		\end{equation}
		As the two-dimensional radial parts of the metrics (\ref{NonextKruskal}) and (\ref{extKruskal}) are conformally flat, so the entropy formula (\ref{eq3.1}) is applicable.
		\item If the distance $L$ between region $A$ and region $B$ is sufficiently small, Then the finite part of the entanglement entropy of massless matter fields can be evaluated by \cite{Hashimoto:2020cas,Casini:2005zv,Casini:2009sr}
		\begin{equation}
			\label{eq3.2}
			S_{\rm{matter}}^{\rm{finite}}=-I(A:B)=-\kappa_{d+1}c\frac{Area}{L^{d-1}} ,
		\end{equation}
		where $c$ is the central charge and $\kappa_{d+1}$ is a dimensionally
		dependent constant. Note that this formula is only valid for the flat spacetime. While we expect the above formula can also be applied in curved spacetime as long as we require that the length scale of the curvature is much larger than the distance $L$.
		\end{itemize}

	\section{The entanglement entropy in non-extremal large $d$ RN-AdS$_{d+1}$ black hole}
	\label{sec.4}
\begin{figure}[htbp]
		\centering
		\subfloat[Without island]{\includegraphics[width=0.42\columnwidth,height=0.6\linewidth]{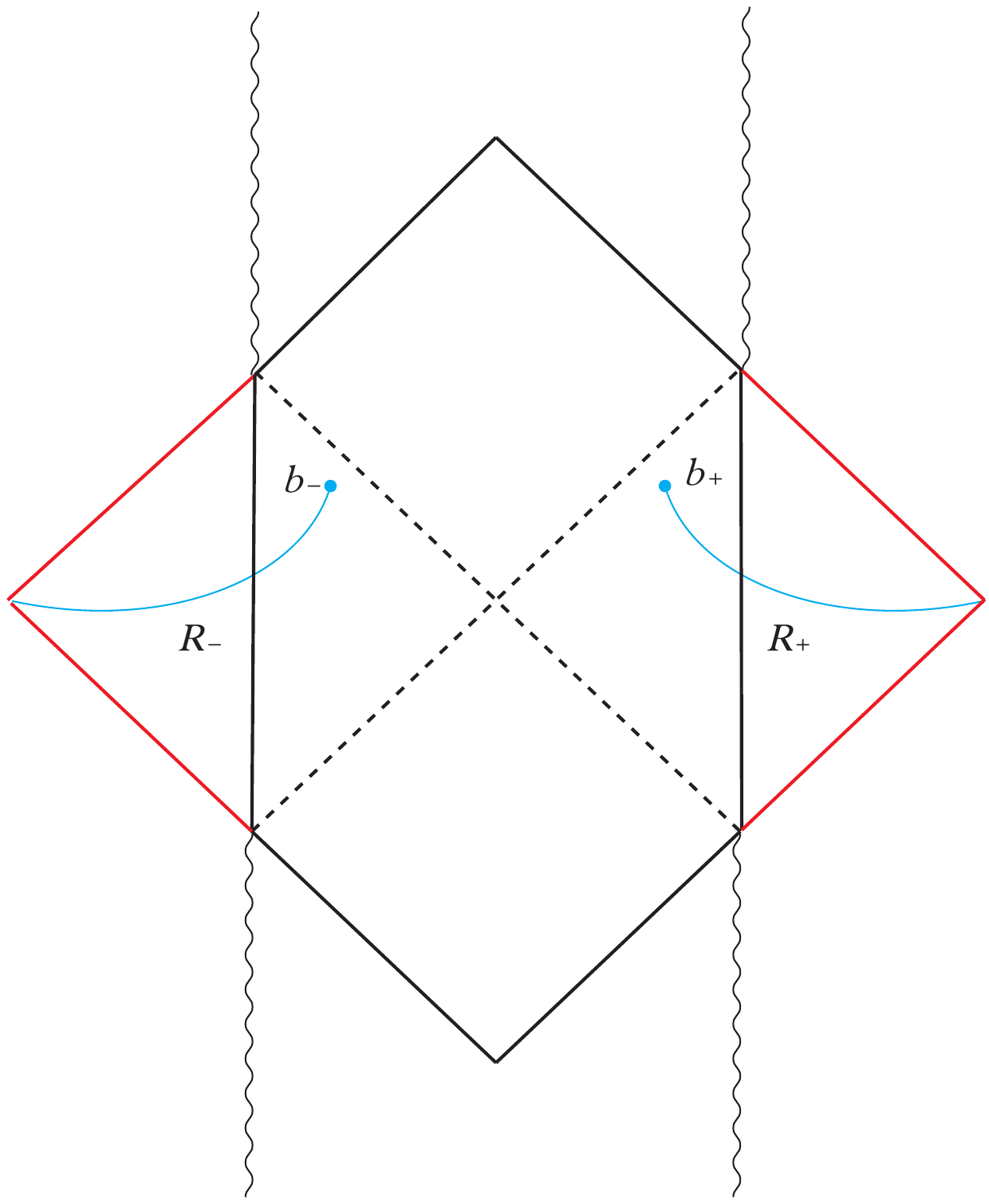}\label{fig.1}}
		\quad
		\subfloat[With island]{\includegraphics[width=0.42\columnwidth,height=0.6\linewidth]{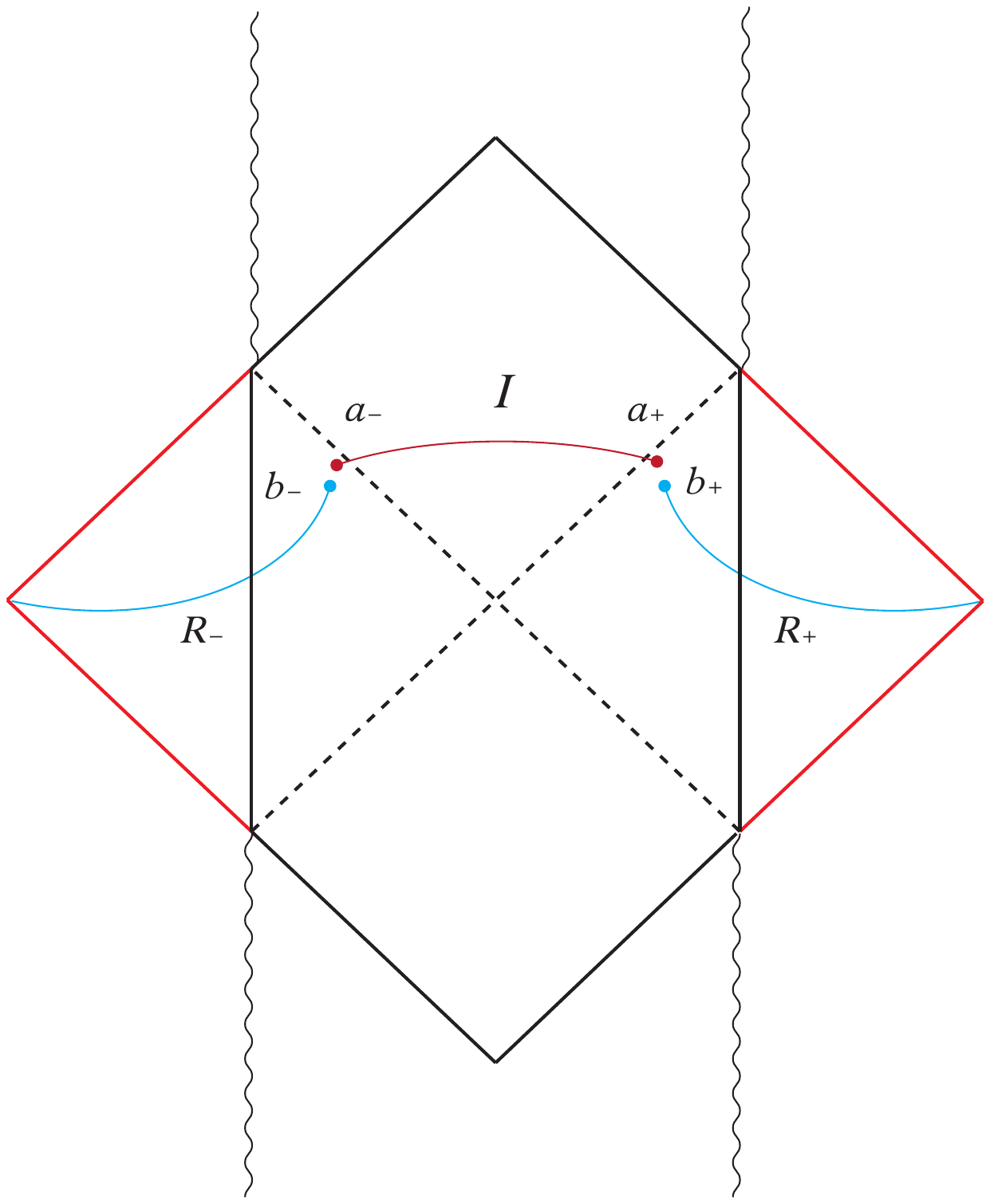}\label{fig.2}}
		
		\caption{The Penrose diagram of the nonextremal RN-AdS$_{d+1}$ black hole coupled to two auxiliary thermal baths. The black part and red triangle represent the black hole and the auxiliary at spacetime, respectively. \textbf{(a)} $R_{\pm}$ are the radiation regions on the right and left wedges, and $b_{\pm}$ are the boundaries of the radiation region $R_{\pm}$. \textbf{(b)} The boundaries of the island are supposed at $a_{\pm}$, and the inner boundaries of radiation regions correspond to $b_{\pm}$.}
		\label{fig.nonexetbh}
	\end{figure}

	In this section, we will calculate the entanglement entropy of radiation in the non-extremal large $d$ RN-AdS$_{d+1}$ black hole. Before discussing the island rule, we first give a specific description to the model we studied. In order to investigate the evaporation process of black hole in AdS spacetime, we expect to couple a bath at the asymptotically AdS boundary of the black hole. Here, we couple two flat thermal bath systems which have no gravitational effect at the boundary of RN-AdS$_{d+1}$ black hole, and make it transparent \cite{Yu:2021rfg,Rocha:2008fe}. For the thermal bath, suppose that the bath is in thermal equilibrium with the black hole. The Penrose diagram of whole spacetime~(RN-AdS$_{d+1}$+bath) is shown in Fig. \ref{fig.nonexetbh}. 
	
	  Furthermore, it is worth noting that we choose the boundary of radiation region $b_{\pm}$ near horizon, which is crucial in our later calculation of the entanglement entropy of Hawking radiation. As we can see, due to the special features of large $d$ geometry near the horizon, the metric of large $d$ RN-AdS black hole behaves like in asymptotically flat spacetime~\cite{Emparan:2013moa,Emparan:2020inr}. Then a natural setup is to choose the boundary of the radiation region $b_{\pm}$ that is to be  near the horizon, which is similar to the previous models \cite{Hashimoto:2020cas,Du:2022vvg}. To explain this, we can focus on the black hole solution $f(r)$, the horizon radius $r_{+}$ is satisfied with $f(r_{+})=0$ and $r_{+}<r_{o}$. At large $d$,
	\begin{equation}
		r_{+}\simeq r_{o}\left(\frac{1+\sqrt{1-\frac{4kQ^{2}}{M^{2}} } }{2k} \right)^{\frac{1}{d-2} }\simeq r_{0}\left(1-\frac{1}{d} \ln \left(1+\frac{r_{0}^{2}}{L^{2}}\right)+O\left(d^{-2}\right)\right),
	\end{equation}
	when taking $d \to \infty$, $r_{+} \to r_{o}$. It implicates that the gravitational effect of the black hole quickly disappears outside horizon $r>r_{o}$ in large $d$ limit. In fact, there is a small region around the horizon on the $r/d$ scale where the gravitational effect of the black hole is still appreciable, more precisely, within the region
	\begin{equation}
	r-r_{o} \lesssim \frac{r_{o}}{d}+O\left(d^{-2}\right),
	\end{equation}
	 i.e., $\rho= \left(\frac{r_{o}}{r}\right)^{d-2}=O\left(d^{0}\right)$. Thus, the gravitational influence of black hole is mainly concentrated in the near horizon and vanishes in the far zone\footnote{The definition of two distinct region in the geometry: near-horizon region: $r-r_{o}\ll r_{o},~\mathrm{i.e.,}~\ln \rho \ll d$, far region: $r-r_{o}\gg \frac{r_{o}}{d}, ~\mathrm{i.e.,}~\ln \rho \gg 1$.} in large $d$ limit. Therefore, it allows us to choose the boundary of the radiation region $b_{\pm}$ to be near the horizon like in asymptotically flat spacetime. And we are able to calculate the entanglement entropy in large $d$ RN-AdS black hole spacetime by using the formulas~(\ref{eq3.1}) and~(\ref{eq3.2}) as in the following sections. 
	
	\subsection{Without island}
	There are two points $b_{\pm}$ corresponding to the boundaries of radiation regions on left wedge $R_{-}$ and right wedge $R_{+}$ (see the Fig. \ref{fig.1}). Here $b_{+}=(t_{b}, \rho _{b})$ and $b_{-}=(-t_{b}+i \beta /2, \rho _{b})$ respectively. Note that for the case without island, it can be inferred from the island formula that the entanglement entropy of Hawking radiation only receives the contribution from the matter fields. If the distance between the boundaries of $R_{-}$ and $R_{+}$ is large, the entanglement entropy of Hawking radiation can be approximated by formula~(\ref{eq3.1})
	\begin{equation}
		S_{R}=S_{matter}^{finite}(R)=-I(R_{+}:R_{-})=\frac{c}{3}\log d(b_{+},d_{-}) ,
	\end{equation}
	then calculating in the Kruskal coordinates (\ref{NonextKruskal}), we have
	\begin{equation}
		\begin{aligned}
			S_{R}&=\frac{c}{6}\log [g^{2}(b)((U(b_{-})-U(b_{+}))(V(b_{+})-V(b_{-})))]  \\
			&=\frac{c}{6}\log [4\frac{k(\rho -\rho _{+})(\rho -\rho _{-})}{\rho ^{2}}\frac{1}{\kappa _{+}^{2}}e^{-4\kappa _{+}\rho _{*}(b)} \cosh ^{2}(\kappa _{+}t_{b}) ] .
		\end{aligned}
	\end{equation}
	At late time, we assume that $t_{b} \gg \rho _{b} > \rho_{+}$, thus
	\begin{equation}
		\label{eq4.3}
		S_{R} \simeq \frac{c}{3}\kappa _{+} t_{b} = \frac{d-2}{r_{o}}\frac{k(\rho_{+}-\rho_{-})}{6\rho_{+}}t_{b} .
	\end{equation}
	We can see that the entanglement entropy of the radiation increases linearly with time at late time and becomes larger than the Bekenstein-Hawking entropy. This clearly does not satisfy the unitary which requires the entanglement entropy to follow the Page curve. However, this problem will be solved once we introduce the contribution of the island after the Page time.
	\subsection{With island}
	
	Now we consider the contribution of the island to the entanglement entropy of Hawking radiation. We will focus on the situation where the inner boundary of the radiation region is near the outer horizon, characterized by $\rho _{b}-\rho _{+}\ll \rho _{+}$. We set the boundaries of the island as $a_{+}=(t_{a}, \rho _{a})$ and $a_{-}=(-t_{a}+i\beta/2, \rho _{a})$ respectively (see the Fig. \ref{fig.2}). Here we will apply the formula (\ref{eq3.2}) to calculate the entanglement entropy of matter fields, as we have set the boundary of the island to be outside and near the outer horizon, namely, $\rho _{a}-\rho _{+}< \rho _{b}-\rho _{+}\ll \rho _{+}$.  By formula~(\ref{eq3.2}), we have
	\begin{equation}
		S_{matter}^{finite}(R \cup I)=-2I(R_{+}:I)=-2\kappa _{d+1}c\frac{Area}{L^{d-1}} ,
	\end{equation}
	the geodesic distance between the boundary of region $I$ and that of region $R$ is~\cite{Hashimoto:2020cas}
	\begin{equation}
		\begin{aligned}
			L&=\int_{\rho_{a}}^{\rho_{b}} \frac{r_{o}}{d-2} \frac{d\rho }{\sqrt[]{k(\rho-\rho_{+})(\rho-\rho_{-})} } \\
			&\simeq 2\frac{r_{o}}{d-2} \frac{1}{\sqrt[]{k(\rho_{+}-\rho_{-})}}(\sqrt{\rho_{b}-\rho _{+}}-\sqrt[]{\rho_{a}-\rho _{+}} ) .
		\end{aligned}
	\end{equation}
	Thus the generalized entropy is
	\begin{equation}
		\label{eq4.6}
		\begin{aligned}
			S_{gen }&=2\frac{Area(\partial I)}{4G_{N}}-2 \kappa _{d+1}c\frac{Area(\partial R)}{L^{d-1}} \\
			&= \frac{\Omega_{d-1}(M\rho _{a})^{\frac{d-1}{d-2} }}{2G_{N}}\\
			&-2\kappa_{d+1}c\frac{\Omega_{d-1}(M\rho _{b})^{\frac{d-1}{d-2}}}{(\sqrt{\rho _{b}-\rho _{+}}-\sqrt[]{\rho _{a}-\rho _{+}} ) ^{d-1}} \frac{(k(\rho _{+}-\rho _{-}))^{\frac{d-1}{2} }}{2^{d-1}} \left(\frac{d-2}{r_{o}} \right)^{d-1} .
		\end{aligned}
	\end{equation}
	The factor 2 is due to the double contributions from the left and right wedges. For convenience, we define new variables $x\equiv \sqrt{\frac{\rho _{a}-\rho _{+}}{\rho _{+}} }$ and $y\equiv \sqrt{\frac{\rho _{b}-\rho _{+}}{\rho _{+}} }$, thus $x<y\ll 1$ because of $\rho _{a}-\rho _{+}< \rho _{b}-\rho _{+}\ll \rho _{+}$. Then eq. (\ref{eq4.6}) becomes
	\begin{equation}
		\label{eq4.7}
		S_{gen}= \frac{\Omega_{d-1}(M\rho _{a})^{\frac{d-1}{d-2} }}{2G_{N}}-2\kappa_{d+1}c\frac{\Omega_{d-1}(M\rho _{b})^{\frac{d-1}{d-2}}}{\left( 1-\frac{x}{y} \right) ^{d-1}} \frac{(k(\rho _{+}-\rho _{-}))^{\frac{d-1}{2} }}{\rho _{+}^{\frac{d-1}{2} }y^{d-1}2^{d-1}} \left(\frac{d-2}{r_{o}} \right)^{d-1} .
	\end{equation}
	According to the island formula, the entanglement entropy is given by the minimal value among all extremal solutions of the generalized entropy. Note that the expression of the generalized entropy by using formula (\ref{eq3.2}) does not explicitly include time. We just take 
	\begin{figure}[htbp]
		\centering
		\includegraphics[width=0.75\columnwidth,height=0.45\linewidth]{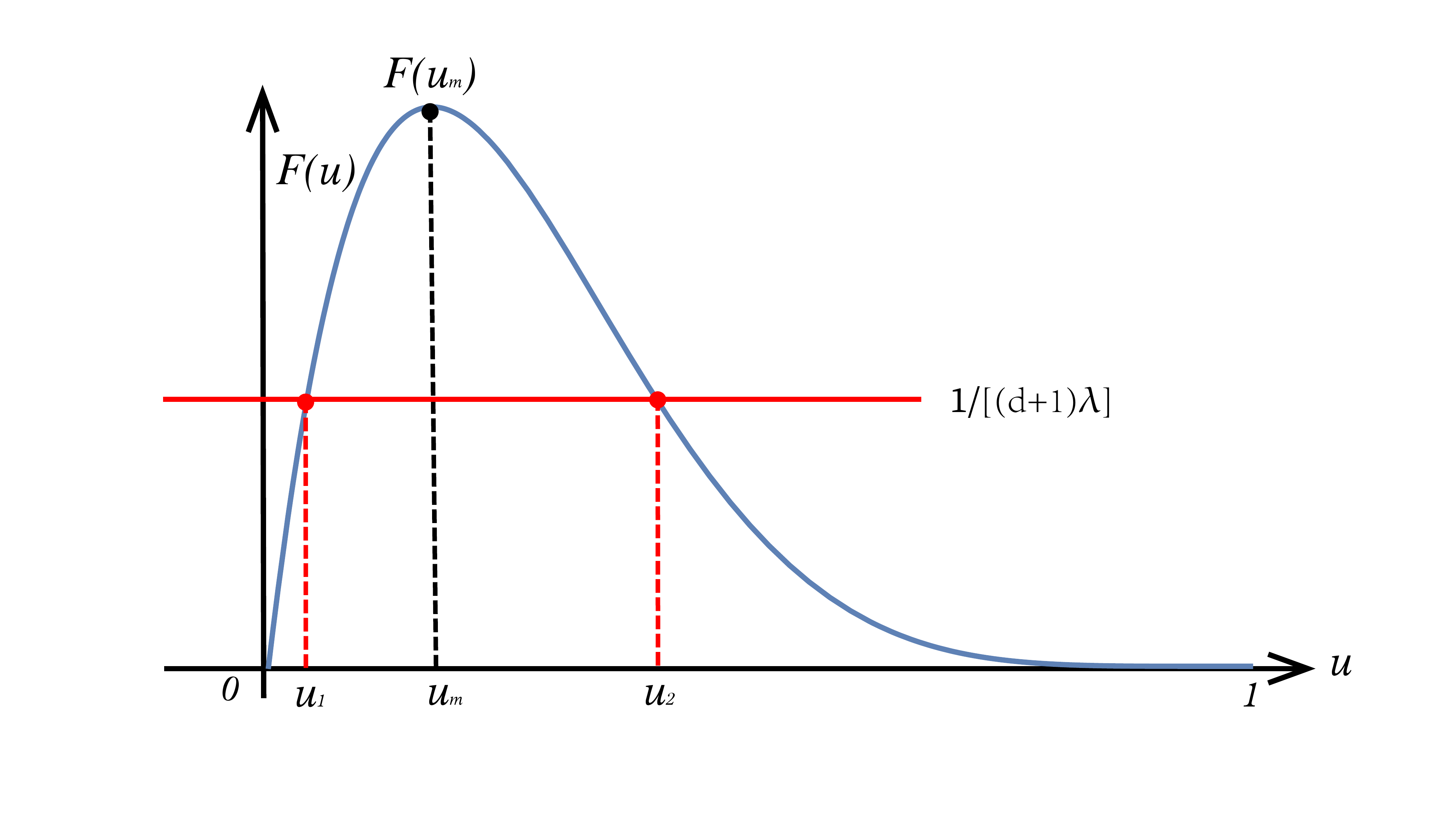}
		\caption{The schematic diagram of function $F(u)$ in the interval $u \in (0, 1) $. A local maximum value of function $F(u)$ is located at $u_{m}$. There are two solutions $u_{1}$ and $u_{2}$ when $F(u_{m}) > 1/[(d+1) \lambda]$. The island solution is just $u_{1}$.}
		\label{fig.3}
	\end{figure}
	the derivative with respect to position $\rho _{a}$ and solve the following equation
	\begin{equation}
		\label{eq4.8}
		\begin{aligned}
			\frac{\partial S_{gen}}{\partial \rho _{a}}&=\frac{d-1}{d-2} \frac{\Omega_{d-1}M^{\frac{d-1}{d-2} }{\rho _{a}}^{\frac{1}{d-2} }}{2G_{N}}\\
			&-(d-1)\kappa_{d+1}c\frac{\Omega_{d-1}(M\rho _{b})^{\frac{d-1}{d-2}}}{\left( 1-\frac{x}{y} \right) ^{d}xy\rho_{+}} \frac{(k(\rho _{+}-\rho _{-}))^{\frac{d-1}{2} }}{\rho _{+}^{\frac{d-1}{2} }y^{d-1}2^{d-1}} \left(\frac{d-2}{r_{o}} \right)^{d-1}=0 ,
		\end{aligned}
	\end{equation}
	which gives
	\begin{equation}
		\label{eq4.9}
		\frac{x}{y}\left(1-\frac{x}{y}\right)^{d}=G_{N}\kappa _{d+1}c\frac{(k(\rho _{+}-\rho_{-} ))^{\frac{d-1}{2} }(d-2)^{d}}{\rho _{+}^{\frac{d-1}{2} }y^{d+1}2^{d-2}r_{o}^{d-1}} .
	\end{equation}

	\subsubsection{The existence of general island solution}
	\label{sec.4.2.1}
	Now let us study the general island solution of eq. (\ref{eq4.9}).
	And we will show that the existence of island solution requires some constraints on the large $d$ RN-AdS$_{d+1}$ black hole. Similar observations in Schwarzschild black hole spacetime have been noticed in Refs. \cite{Arefeva:2021kfx,Matsuo:2020ypv,Du:2022vvg,He:2021mst}. By defining new variables as
	\begin{equation}
		u\equiv \frac{x}{y} \in (0, 1)\quad \text{and} \quad\lambda \equiv \frac{\rho _{+}^{\frac{d-1}{2} }y^{d+1}2^{d-2}r^{d-1}_{o}}{c\kappa _{d+1}G_{N}(k(\rho _{+}-\rho _{-}))^{\frac{d-1}{2} }(d-2)^{d}(d+1)}.
	\end{equation}
	where $ x< y\ll 1 $. Eq. (\ref{eq4.9}) becomes
	\begin{equation}
		\label{eq4.11}
		u(1-u)^{d}=\frac{1}{(d+1)\lambda } .
	\end{equation}
	Note there exists a local maximum value of function $F(u)$, that is
	\begin{equation}
		\quad F(u_{m})=\frac{d^{d}}{(d+1)^{d+1}}\quad \text{at}\quad  u_{m}=\frac{1}{d+1}.
	\end{equation}
	\begin{figure}[htbp]
		\centering
		\subfloat[$\lambda > e$]{\includegraphics[width=0.48\columnwidth,height=0.4\linewidth]{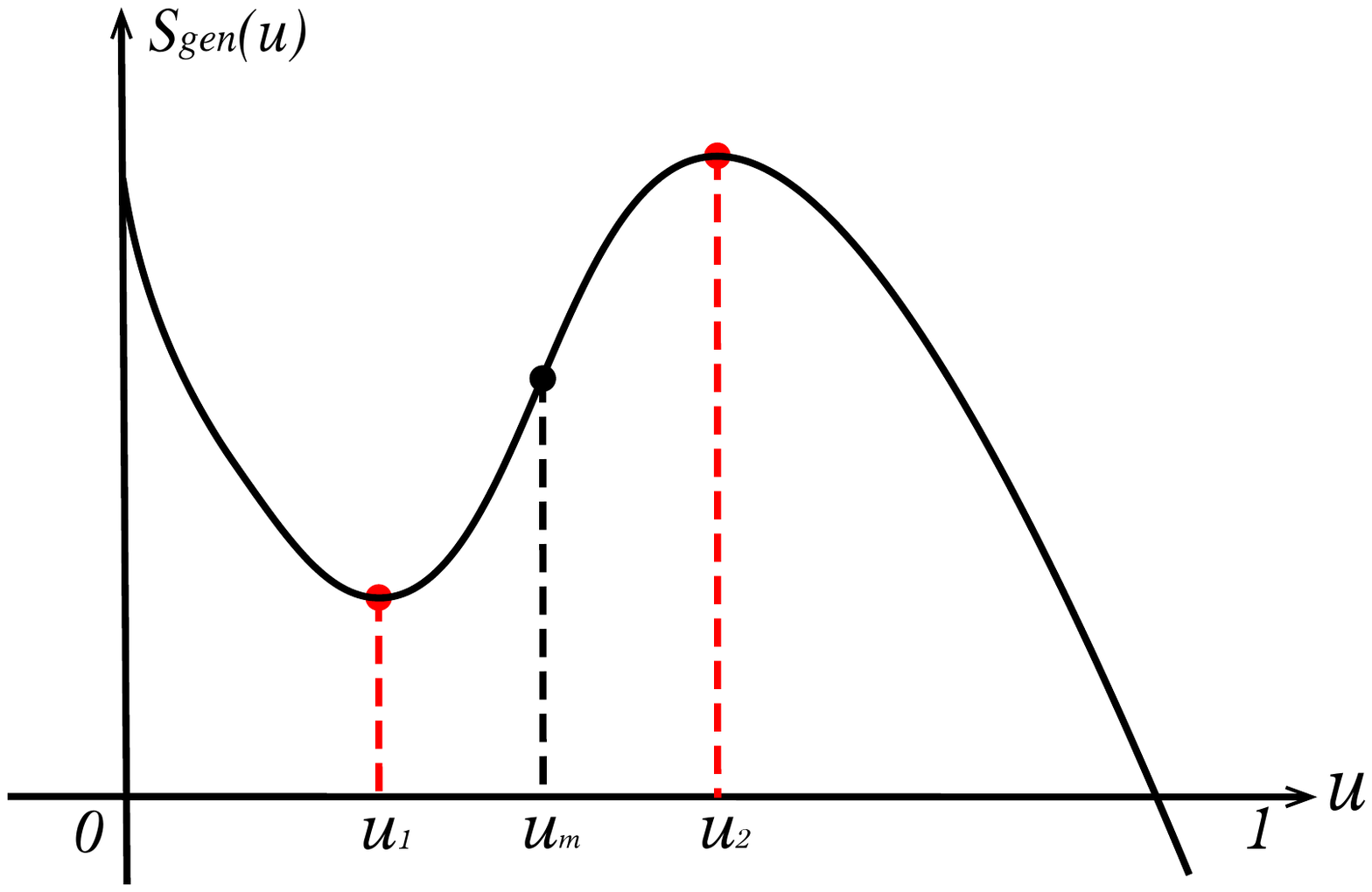}}
		\quad
		\subfloat[$\lambda \le e $]{\includegraphics[width=0.48\columnwidth,height=0.4\linewidth]{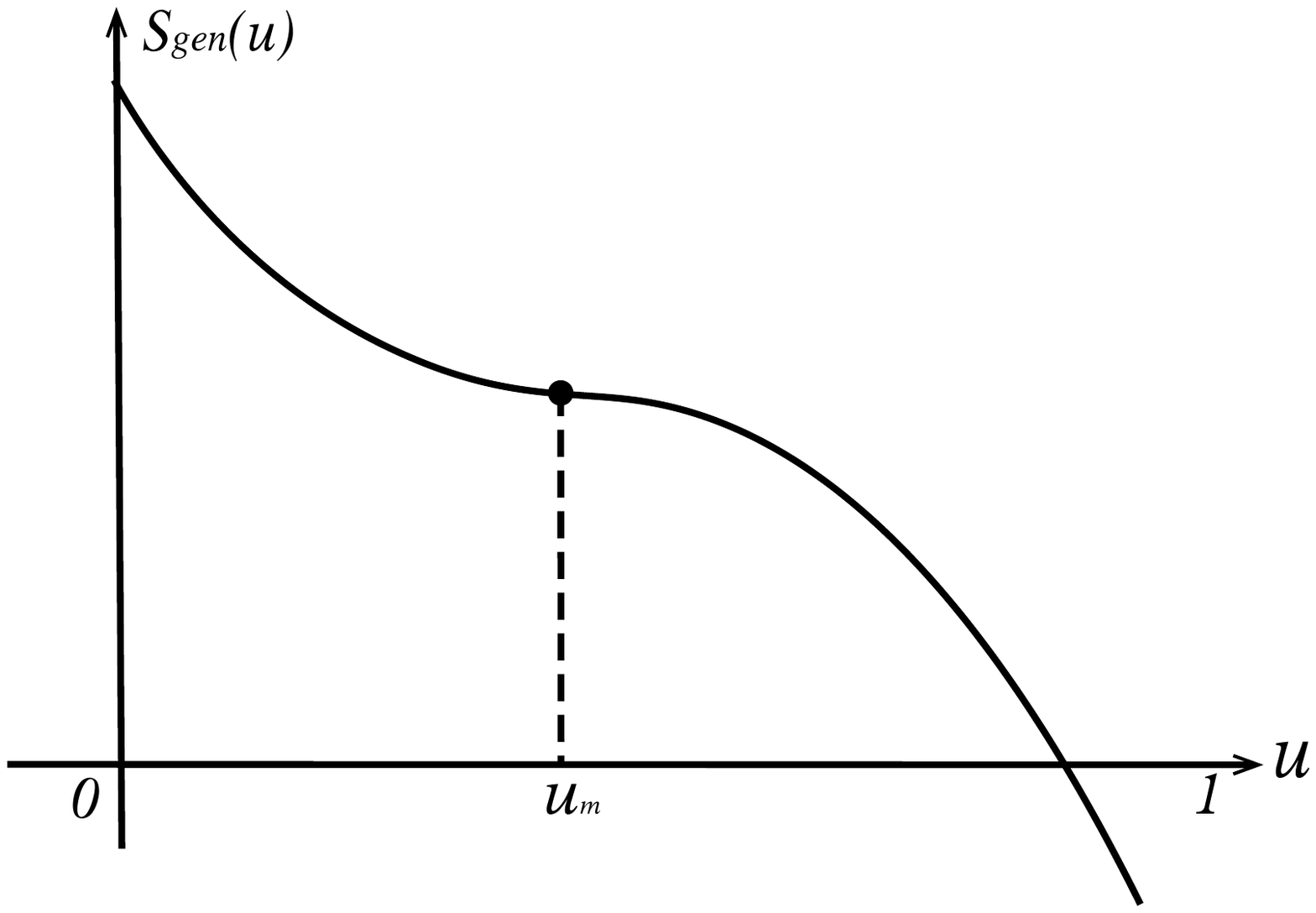}}
		
		\caption{The entropy curve of $S_{gen}$ with $u$. \textbf{(a)} When $\lambda >e $, the generalized entropy $S_{gen}$ reaches its local minimum and local maximum at $u_{1}$ and $u_{2}$ respectively. \textbf{(b)} If $\lambda \le e $, there would be no solutions.}
		\label{fig.4}
	\end{figure}
	And function $F(u) = u(1-u)^{d}$ monotonically increases with $u$ in the interval $(0,u_{m})$ and monotonically decreases in the interval $\left [ u_{m},1 \right ) $ (see the Fig. \ref{fig.3}). Obviously, the eq. (\ref{eq4.11}) exists a island solution only if $\frac{1}{(d+1)\lambda} < F(u_{m})$. Therefore, in the large  $d$ limit, we obtain
	\begin{equation}\label{constraint}
		\lambda > \left(1+\frac{1}{d}\right)^d \to e.
	\end{equation}
	As shown in Fig. \ref{fig.4}, if the constraint $\lambda > e$ is satisfied, there would exist two solutions $u_{1}$ and $u_{2}$ ($0<u_{1} < u_{m} < u_{2}<1$). The local minimum and maximum values of the generalized entropy are located at $u=u_{1}$ and $u=u_{2}$ respectively. Thus the island solution is exactly given by $u=u_{1}$. When $\lambda = e$, points $u_{1}$ coincide with $u_{2}$ at $u_{m}$, the generalized entropy $S_{gen}$ will monotonically decrease with $u$ in the interval $(0,1)$. Thus there would be no local minimum value of $S_{gen}$ and no nontrivial island solution can be found, similarly for the case $\lambda < e$. Physically speaking, the existence of the island puts a constraint on the large dimensional RN-AdS$_{d+1}$ black hole. As according to the island formula, we need to make sure the existence of the island in order to save the unitarity in our case, and this makes the constraint (\ref{constraint}) meaningful.
	
	\subsubsection{Two specific analytical island solutions}
	\label{sec.4.2.2}
	Note that we can not give an analytical expression of island solution for general case $x< y \ll 1$. In this section, we  will give two analytical island solutions for the special cases $x \ll {y}/{d}$ and $x \sim {y}/{d}$ in the large $d$ limit.
	
	First, we consider a more special case satisfying $x \ll {y}/{d}$ with $d\to \infty$. Starting from eq. (\ref{eq4.8}), in which $\left(1-\frac{x}{y}\right)^{-d}$ can be expanded as
	\begin{equation}
		\label{eq4.15}
		\left(1-\frac{x}{y} \right)^{-d}\simeq 1+d\frac{x}{y}+ \mathcal{O}((d\frac{x}{y} )^{2}) .
	\end{equation}
	So it allows us to ignore higher-order terms of eq. (\ref{eq4.15}) in large $d$ limit for special case $x \ll y/d$. Note that in finite dimension case, the condition $x \ll y$ would be enough to drop the higher-order terms of eq. (\ref{eq4.15}). But in the large dimension case, the condition $x \ll y$ is not enough and we need further assume $x \ll y/d$. Substituting eq. (\ref{eq4.15}) into eq. (\ref{eq4.8}) and omitting the higher-order terms, we find an island solution as
	\begin{equation}
		\label{eq4.18}
		x=\frac{y}{\frac{y^{d+1}r_{o}^{d-1}2^{d-2}}{(d-2)^{d}\kappa _{d+1}cG_{N}}\left(\frac{\rho_{+} }{k(\rho _{+}-\rho_{-})}\right)^{\frac{d-1}{2} }  -d}=\frac{y}{(d+1)\lambda-d},
	\end{equation}
	which is a special island solution that is valid for $x \ll y/d$ with $d\rightarrow\infty$.
	
\begin{figure}[htbp]
		\centering
		\includegraphics[width=0.75\columnwidth,height=0.45\linewidth]{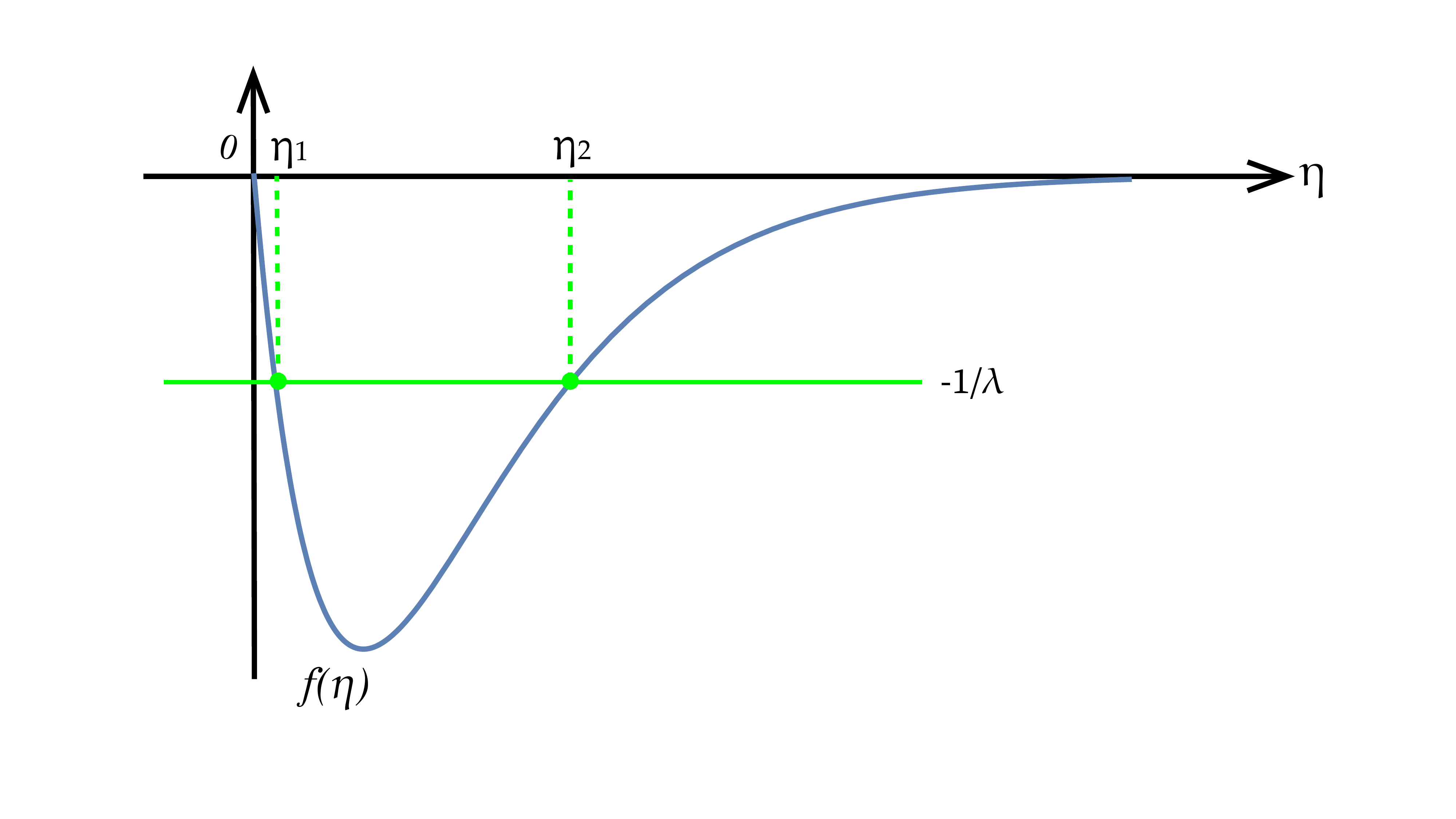}
		\caption{The schematic diagram of the function $f(\eta)= -\eta e^{-\eta}$. There are two solutions $\eta_{1}$ and $\eta_{2}$ of eq. (\ref{eq4.20}). The island solution is just $\eta_{1}$.}
		\label{fig.5}
	\end{figure}

	Now we try to give another special analytical solution in the large $d$ limit. We assume $d\frac{x}{y}=\eta \sim O(1)$, so that $\lim_{d \to \infty}\left(1-\frac{\eta }{d} \right)^{d}=e^{-\eta } $. Then from eq. (\ref{eq4.11}), we get
	\begin{equation}
		\label{eq4.20}
		-\eta e^{-\eta}=-\frac{1}{\lambda} .
	\end{equation}
	The solution of the above equation can be expressed as the Lambert $W$ function (or the product logarithmic function). It can be seen in the Fig. (\ref{fig.5}) that there are two real number solutions $\eta_{1}=-W\left(-\frac{1}{\lambda}\right)$ and $\eta_{2}=-W_{-1}\left(-\frac{1}{\lambda}\right)$, where $-\frac{1}{\lambda} \in (-\frac{1}{e}, 0)$. The island solution is just given by $\eta_{1}=-W\left(-\frac{1}{\lambda}\right)$, that is
	\begin{equation}
		\label{eq4.21}
		x=\frac{y}{d}\eta_{1}=-\frac{y}{d}W\left(-\frac{1}{\lambda}\right),
	\end{equation}
	which is a special island solution that is valid for $x \sim y/d$ with $d\rightarrow\infty$.
	
	\subsubsection{The entanglement entropy with island}
	In previous sections, we have confirmed the existence of the island solution for the general case $x< y \ll 1$. Now let us evaluate the entanglement entropy of Hawking radiation when the island exists. For the special case with $x \ll y/d $ and $d\rightarrow\infty$, we note that the solution (\ref{eq4.18}) should satisfy the constraint $\lambda \gg 1$. Thus
	\begin{equation}
		\label{eq4.24}
		x=\frac{y}{(d+1)\lambda-d}\simeq \frac{y}{(d+1)\lambda}.
	\end{equation}
	Then by plugging the solution (\ref{eq4.24}) back into the formula (\ref{eq4.7}), the entanglement entropy in this special case is given by
	\begin{equation}
		\label{eq4.25}
		\begin{aligned}
			S_{R}\simeq &\frac{\Omega_{d-1}(M\rho_{+})^{\frac{d-1}{d-2} }}{2G_{N}}\\
			&-\kappa_{d+1}c\frac{\Omega_{d-1}(M\rho_{+})^{\frac{d-1}{d-2}}}{y^{d-1}} \frac{(k(\rho _{+}-\rho _{-}))^{\frac{d-1}{2} }}{\rho _{+}^{\frac{d-1}{2} }2^{d-2}} \left(\frac{d-2}{r_{o}} \right)^{d-1}\left(1+(d-1)\frac{x}{y}\right)\\
\simeq &2S_{BH}\left(1-\frac{2y^{2}}{(d+1)(d-2)\lambda } \right)
\simeq 2S_{BH} ,
		\end{aligned}
	\end{equation}
	where we used $\lambda \gg 1$ and $y/d \ll y \ll 1$. Therefore, for this special case, the entanglement entropy of Hawking radiation $S_{R}$ is approximately equal to $2S_{BH}$.
	
For another special case with $d\frac{x}{y}=\eta \sim O(1)$, from eq. (\ref{eq4.20}) and eq. (\ref{eq4.21}) we have the following expression
	\begin{equation}
		\label{sc2}
		\left(1-\frac{x}{y}\right)^{d-1} \simeq e^{-\eta_{1}}=\frac{1}{\lambda \eta_{1}}.
	\end{equation}
	Substituting eq. (\ref{eq4.7}) into eq. (\ref{sc2}) , we obtain
	\begin{equation}
		\begin{aligned}
			S_{R} &\simeq2S_{BH}\left(1-\frac{2y^{2}}{d^{2}\lambda }\frac{1}{(1-\frac{x}{y} )^{d-1}} \right ) \\
			&\simeq 2S_{BH}\left(1-\frac{2y^{2}}{d^{2}}\eta _{1} \right)
			\simeq 2S_{BH},
		\end{aligned}
	\end{equation}
where we have used the solution $\eta_{1}\sim O(1)$ and $y/d \ll y \ll 1$, Therefore, we obtained the entanglement entropy of Hawking radiation which is approximately equal to $2S_{BH}$ for two special cases.
	
	Moreover, it can be shown that for the general case $x < y \ll 1$, one still has $S_{R} \simeq 2S_{BH}$. Since from eq. (\ref{eq4.7}), we have
	\begin{equation}
		S_{R} \simeq 2S_{BH}\left(1-\frac{2y^2}{(d+1)(d-2)\lambda}\left(1-\frac{x}{y}\right)^{1-d}\right) <2S_{BH}.
	\end{equation}
	For the general case $x < y \ll 1$, the island solution satisfies $x/y = u_{1} < u_{m}={1}/(d+1) $ under the constraint $\lambda >e $. Then we have
	\begin{equation}\label{eq4.2.1}
		S_{R} > 2S_{BH}\left(1-\frac{2y^2}{(d+1)(d-2)\lambda}(1-\frac{1}{d+1})^{1-d}\right)=2S_{BH}\left(1-\frac{2y^2e}{(d+1)(d-2)\lambda}\right) \simeq 2S_{BH} .
	\end{equation}
	
	\begin{figure}[htbp]
		\centering
		\subfloat[Without island]{\includegraphics[width=0.42\columnwidth,height=0.52\linewidth]{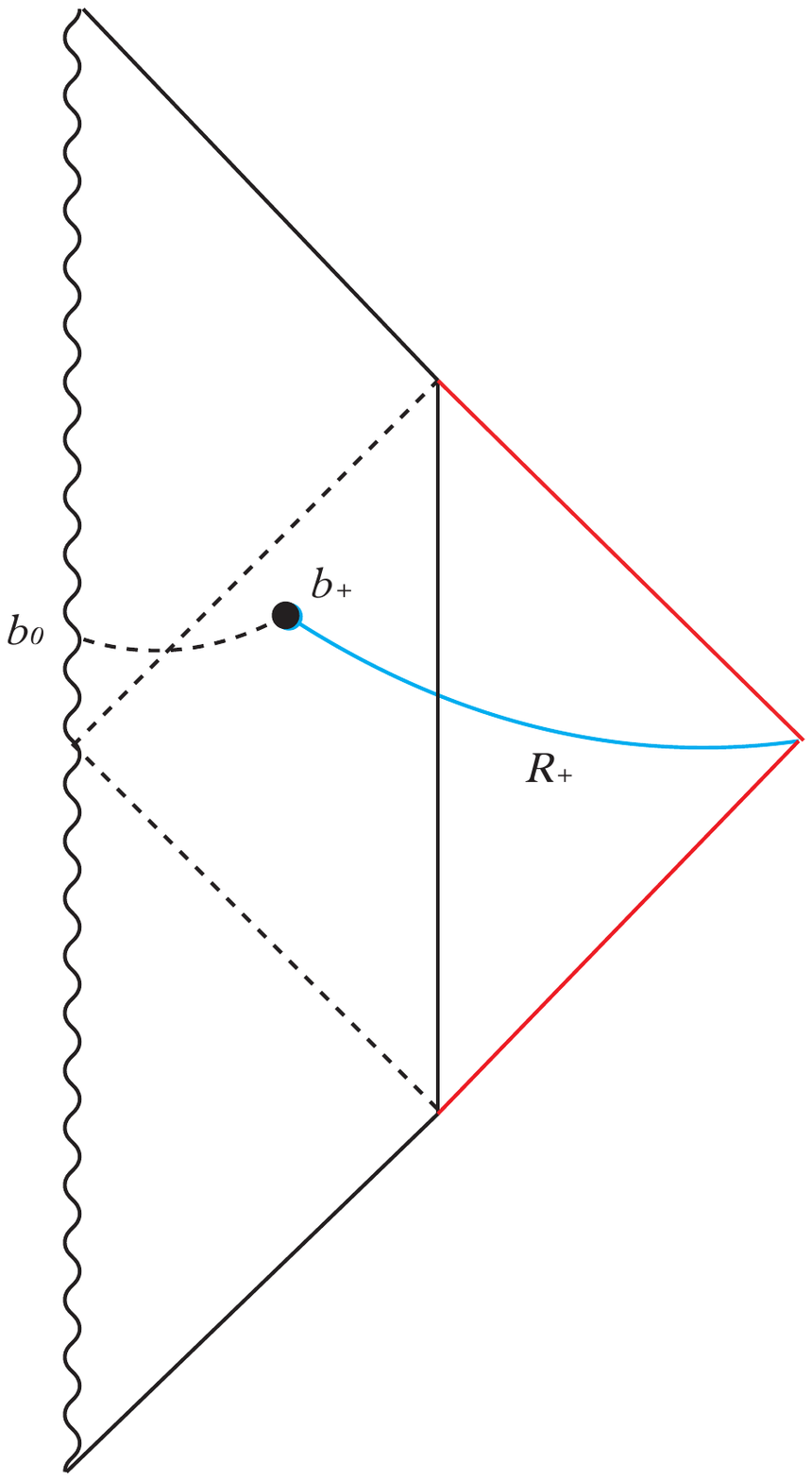}\label{fig.6.1}}
		\quad
		\subfloat[With island]{\includegraphics[width=0.42\columnwidth,height=0.52\linewidth]{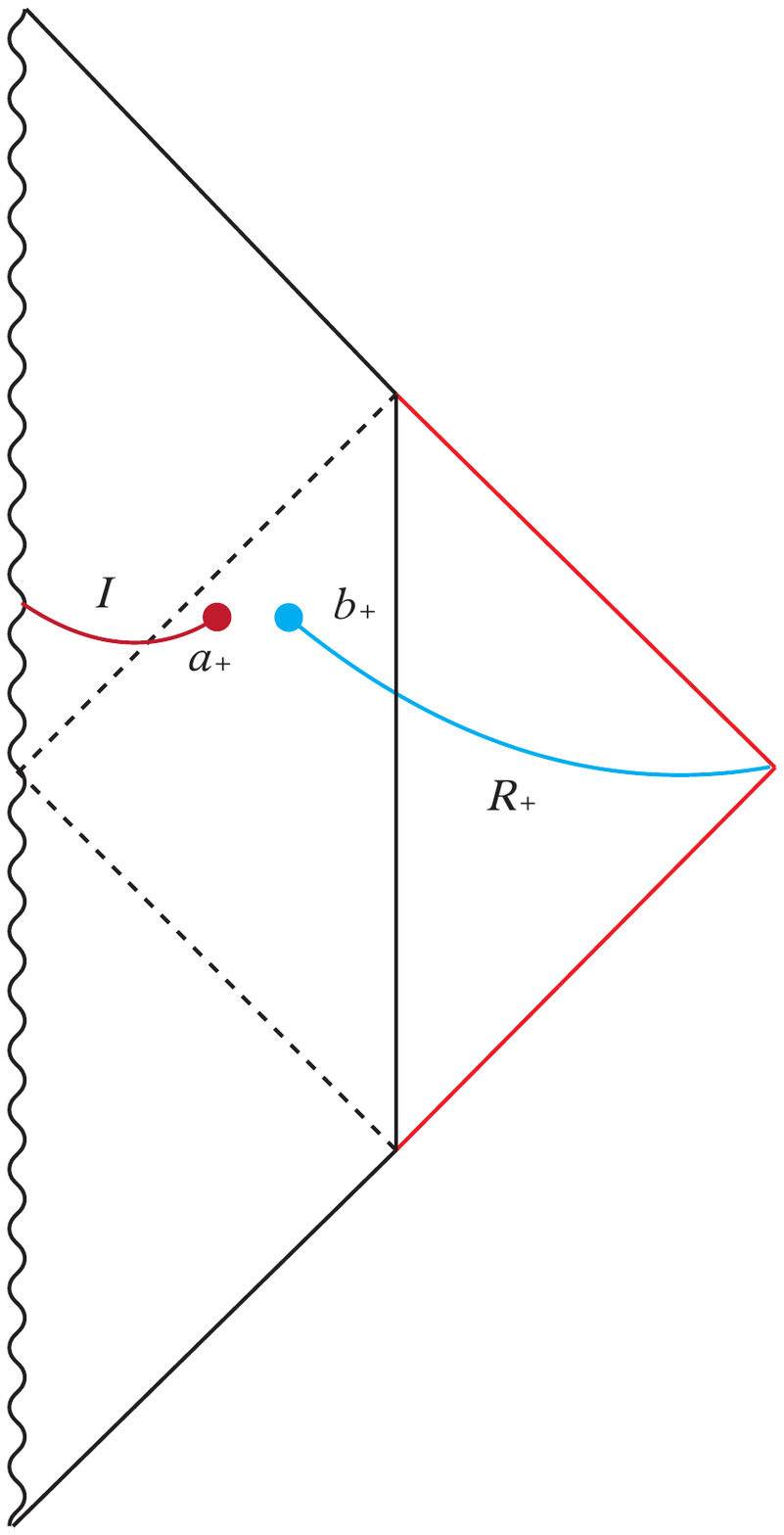}\label{fig.6.2}}
		
		\caption{The Penrose diagram of the extremal RN-AdS$_{d+1}$ black hole coupled to an auxiliary thermal bath. The black part and red triangle represent the black hole and the auxiliary at spacetime, respectively. \textbf{(a)}~$b_{+}$ is the boundary surface of the radiation region $R_{+}$ and $b_{0}$ is the singularity with $\rho =0$. \textbf{(b)}~The island region extends from $\rho =0$ to $\rho=\rho _{a}$.}
	\end{figure}
	Thus we obtain $S_{R}\simeq 2S_{BH}$ for general case $x < y \ll 1$. The leading term is given by double Bekenstein-Hawking entropy, which comes from the boundary area term of the island. The subleading term has been ignored, which reflects the contribution from the quantum effects of matter fields.

\section{The entanglement entropy in  extremal large $d$ RN-AdS$_{d+1}$ black hole}
\label{sec.5}
In this section, we consider the extremal large $d$ RN-AdS$_{d+1}$ black hole. In the same way, we attach an auxiliary bath to the AdS boundary of extremal large $d$ RN-AdS$_{d+1}$ black hole. In Refs.~\cite{Kim:2021gzd,Ahn:2021chg}, the authors argued that one can not calculate the entanglement entropy of the extremal black hole by taking the extremal limit from the entanglement entropy of the non-extremal black hole, because the Penrose diagram of the extremal black hole is not a continuous limit of the non-extremal case, one should start from the extremal setup. And we will start from the Penrose diagram of the extremal black hole to calculate the entanglement entropy in the extremal large $d$ RN-AdS$_{d+1}$ black hole.

	\subsection{Without island}
	As showed in the Fig. \ref{fig.6.1}, the Cauchy surface including $b_{+}=(t_{b}, \rho_{b})$ touches the singularity at $b_{0}=(t_{b}, 0)$. By using eq. (\ref{eq3.1}), the entanglement entropy of Hawking radiation is given by
	\begin{equation}
		\label{eq5.1}
		\begin{aligned}
			S_{R}&=\lim_{\kappa \to 0} \frac{c}{3} \log d(b_{+},b_{0})\\
			&=\lim_{\kappa \to 0} \frac{c}{6} \log \big[ w(\rho_{b})w(0)(U(b_{0})-U(b+))(V(b_{+})-V(b_{0}))) \big]\\
			&=\frac{c}{12}\log \left [f(0)f(\rho_{b})(\rho _{\ast }(\rho_{b})-\rho _{\ast }(0))^{2} \right ] ,
		\end{aligned}
	\end{equation}
	where $\rho_{*}$ is defined in Sec. \ref{sec.2.2}, and $f(\rho)=k(\rho-\rho_{h})^{2}/\rho^{2}$. We can find that $f(0)$ is singular for the extremal large dimensional RN-AdS$_{d+1}$ black hole and the entanglement entropy is divergent at $\rho = 0$. It means that we can not give a well-behaved entanglement entropy of Hawking radiation for the extremal case. This problem also was noticed in Refs. \cite{Kim:2021gzd,Ahn:2021chg,HosseiniMansoori:2022hok}. But we can still give the entanglement entropy of Hawking radiation when the island exists in the extremal case.

	\subsection{With island}
	In the presence of the island, we set the boundary of the island as $a_{+}=(t_{a}, \rho_{a})$. Similarly, we consider the situation where $\rho_{b}-\rho_{h} \ll \rho_{h}$. We assume that the boundary of the island is outside and near the horizon, thus $\rho_{a}-\rho_{h}< \rho_{b}-\rho_{h} \ll \rho_{h}$ (see the Fig. \ref{fig.6.2}), we still use the formula (\ref{eq3.2}) for analysis. Before calculating the entanglement entropy, let us first give the geodesic distance between $a_{+}$ and $b_{+}$:
	\begin{equation}
		L=\int_{\rho_{a}}^{\rho_{b}}\frac{r_{o}}{d-2} \frac{d\rho }{k^{\frac{1}{2}}(\rho -\rho _{h})}=\frac{r_{o}}{k^{\frac{1}{2}}(d-2)}\log \left(\frac{\rho_{b}-\rho_{h}}{\rho_{a}-\rho_{h}} \right) .
	\end{equation}
	By using eq. (\ref{eq3.2}), the generalized entropy is
	\begin{equation}
		\begin{aligned}
			S_{gen}&=\frac{\Omega _{d-1}(M\rho _{a})^{\frac{d-1}{d-2} }}{4G_{N}} -\kappa _{d+1}c\frac{\Omega _{d-1}(M\rho _{b})^{\frac{d-1}{d-2} }}{L^{d-1}} \\
			&=\frac{\Omega _{d-1}(M\rho _{a})^{\frac{d-1}{d-2} }}{4G_{N}} -\kappa _{d+1}c\frac{\Omega _{d-1}(M\rho _{b})^{\frac{d-1}{d-2} }}{\left(\log (\frac{\rho_{b}-\rho_{h}}{\rho_{a}-\rho_{h}} )\right)^{d-1}}\frac{k^{\frac{d-1 }{2}}(d-2)^{d-1} }{r_{o}^{d-1} } .
		\end{aligned}
	\end{equation}
	We still adopt the definitions $x\equiv \sqrt{\frac{\rho _{a}-\rho _{+}}{\rho _{+}} }$ and $y\equiv \sqrt{\frac{\rho _{b}-\rho _{+}}{\rho _{+}} }$, thus $x < y \ll 1$. The generalized entropy becomes
	\begin{equation}
		\label{eq5.4}
		S_{gen}=\frac{\Omega _{d-1}(M\rho _{a})^{\frac{d-1}{d-2} }}{4G_{N}} -\kappa _{d+1}c\frac{\Omega _{d-1}(M\rho _{b})^{\frac{d-1}{d-2} }}{\left(\log \frac{y^2}{x^2} \right)^{d-1}}\frac{k^{\frac{d-1 }{2}}(d-2)^{d-1} }{r_{o}^{d-1} } .
	\end{equation}
	Then from the equation $\frac{\partial S_{gen}}{\partial \rho _{a}}=0$, we obtain
	\begin{equation}
		\label{eq5.5}
		\frac{x^2}{y^2}\left (\log \frac{y^2}{x^2}\right )^{d}=4\kappa _{d+1}cG_{N}\frac{(d-2)^{d}k^{\frac{d-1 }{2}}}{r_{o}^{d-1} y^2 } .
	\end{equation}
	Defining $ z \equiv x^2/y^2 \in (0, 1)$, the above equation becomes
	\begin{equation}
		\label{eq5.6}
		z \left(\log \frac{1}{z} \right)^{d}=4\kappa _{d+1}cG_{N}\frac{(d-2)^{d}k^{\frac{d-1 }{2}}}{r_{o}^{d-1} y^2 }\equiv F_{0} .
	\end{equation}
	\begin{figure}[htbp]
		\centering
		\includegraphics[width=0.75\columnwidth,height=0.45\linewidth]{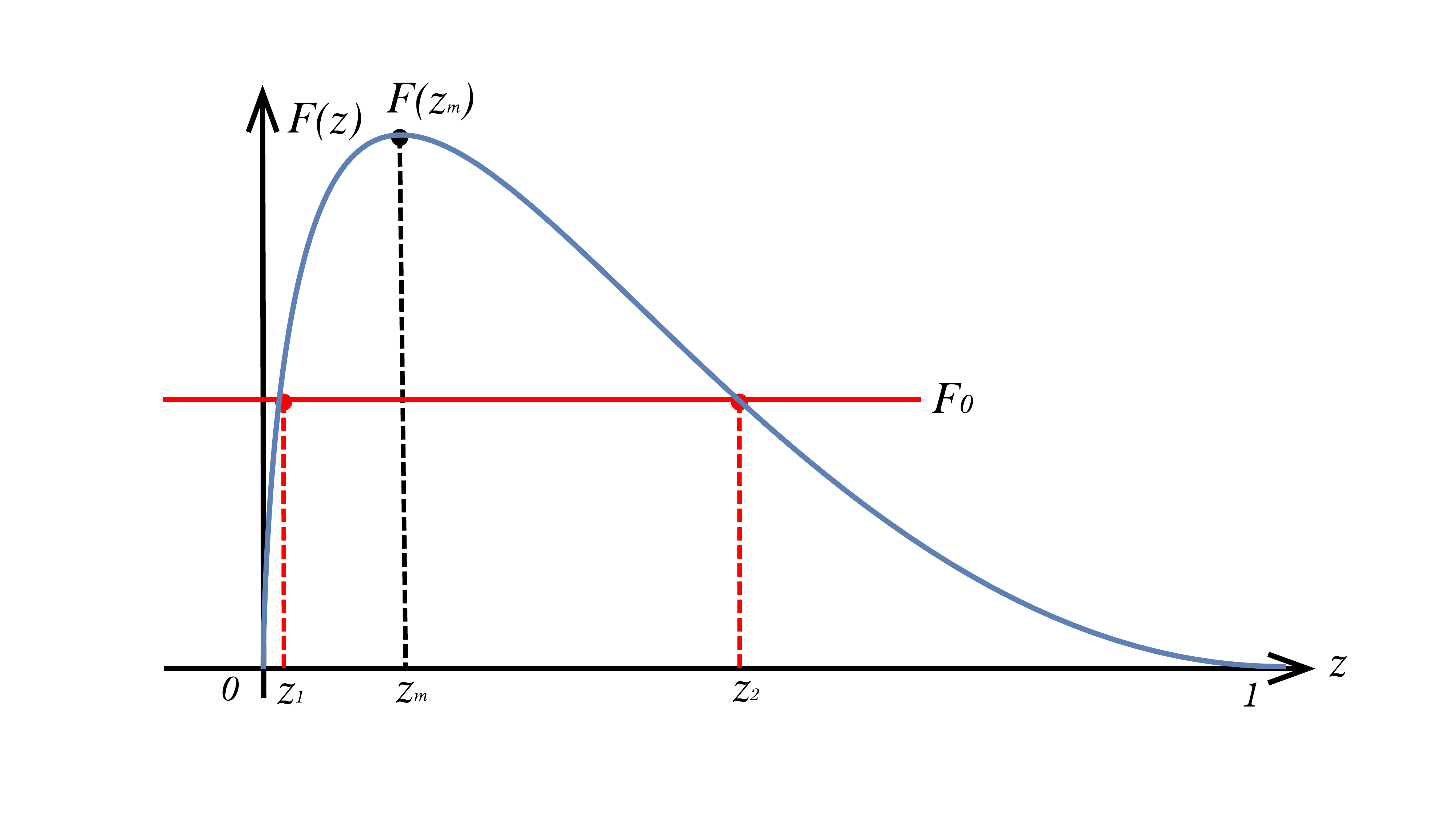}
		\caption{The schematic diagram of function $F(z)$ in the interval $z \in (0,1) $. The maximal value $F(z_{m})$ is located at $z_{m}$. There are two solutions $z_{1}$ and $z_{2}$ when $F(z_{m}) > F_{0}$, and the island solution is given by $z_{1}$.}
		\label{fig.7}
	\end{figure}
	The function $F(z)=z (\log z^{-1})^{d}$ monotonically increases with $z$ in the interval (0, $z_{m}$) and monotonically decreases with $z$ in the interval ($z_{m}$, 1), and the local maximum value of function $F(z)$ is located at $z_{m}$ (see the Fig. \ref{fig.7}), that is
	\begin{equation}
		F(z_{m})= \frac{d^d}{e^d} \quad \text{at} \quad  z_{m}= e^{-d}.
	\end{equation}
	Therefore, the existence of the island solution of eq. (\ref{eq5.6}) requires the constraint
	\begin{equation}
		\label{eq5.8}
		\frac{d^d}{e^d}   > F_{0} .
	\end{equation}
	If the constraint is satisfied, there exists two solutions $z_{1}$ and $z_{2}$ ($z_{1} < z_{m} < z_{2}$). The generalized entropy $S_{gen}$ reaches its local minimum and local maximum at $z_{1}$ and $z_{2}$ respectively, thus the island solution is given by $z_{1}$. If the constraint (\ref{eq5.8}) is violated, there would be no nontrivial island solution in this case. This constraint obtained in the extremal case is different from the constraint (\ref{constraint}) in the non-extremal case.
	
	We have shown the existence of the island solution for $x < y \ll 1$ in extremal case, which requires the constraint (\ref{eq5.8}), though we do not give the exact expression of the island solution in this case. Now we consider the corresponding entanglement entropy of the Hawking radiation. Note that the island solution satisfies $x^2/y^2=z_{1} < z_{m}=e^{-d}$, by plugging the
solution back to eq. (\ref{eq5.4}), we can get the entanglement entropy
	\begin{equation}
		\label{eq5.14}
		\begin{aligned}
			S_{R}&=\frac{\Omega _{d-1}(M\rho _{a})^{\frac{d-1}{d-2} }}{4G_{N}} -\kappa _{d+1}c\frac{\Omega _{d-1}(M\rho _{b})^{\frac{d-1}{d-2} }}{\left(\log \frac{y^2}{x^2} \right)^{d-1}}\frac{k^{\frac{d-1 }{2}}(d-2)^{d-1} }{r_{o}^{d-1} }\\
			&> \frac{\Omega _{d-1}(M\rho _{a})^{\frac{d-1}{d-2} }}{4G_{N}} -\kappa _{d+1}c\frac{\Omega _{d-1}(M\rho _{b})^{\frac{d-1}{d-2} }}{d^{d-1}}\frac{k^{\frac{d-1 }{2}}(d-2)^{d-1} }{r_{o}^{d-1} }\\
			&> S_{BH}\left(1-4\kappa_{d+1} c G_{N}\frac{k^{\frac{d-1}{2}}(d-2)^{d-1}}{d^{d-1}r_{o}^{d-1}}\right) \\
			&> S_{BH}\left(1-\frac{y^2}{e^{d}}\right) \simeq S_{BH},
		\end{aligned}
	\end{equation}
	where $y \ll 1$. We have utilized the following relation from the constraint (\ref{eq5.8}), i.e.
	\begin{equation}
		4\kappa_{d+1} c G_{N}\frac{k^{\frac{d-1}{2}}(d-2)^{d-1}}{d^{d-1}r_{o}^{d-1}}<\frac{d\cdot y^2}{e^{d}(d-2)}<\frac{y^2}{e^{d}}.
	\end{equation}
	At the same time, we have $S_{R} < S_{BH}$, so the value of $S_{R}$ is approximately equal to $S_{BH}$.
	
	In summary, we find that the entanglement entropy is equal to Bekenstein-Hawking entropy for the extremal case. This result is the same as in Ref. \cite{Kim:2021gzd,Ahn:2021chg}, in which the authors consider the situation where the boundary of radiation region is far from the horizon. While we focus on the situation where the boundary of radiation region is taken to be near the horizon in large dimensional RN-AdS$_{d+1}$ black hole, and we mainly calculate the entanglement entropy with island by using the formula (\ref{eq3.2}).

	\section{The constraints in the presence of island}
	\label{sec.6}
	
	In Sec. \ref{sec.4} and Sec. \ref{sec.5}, we have studied the island in non-extremal and extremal large $d$ RN-AdS$_{d+1}$ black hole. In order to ensure the existence of the island to save the unitarity in our cases, the constraints (\ref{constraint}) and (\ref{eq5.8}) should be satisfied. In this section, we would like to analyze these constraints on the large $d$ RN-AdS$_{d+1}$ black hole in more detail.
	
	In the non-extremal case, the constraint (\ref{constraint}) should be satisfied, i.e.
	\begin{equation}
		\label{constraint1}
		\lambda = \frac{\rho _{+}^{\frac{d-1}{2} }y^{d+1}2^{d-2}r^{d-1}_{o}}{c\kappa _{d+1}\ell_{p}^{d-1}(k(\rho _{+}-\rho _{-}))^{\frac{d-1}{2} }(d-2)^{d}(d+1)}>e,
	\end{equation}
	where $G_{N}=\ell_{p}^{d-1}$. Taking the large $d$  limit and utilizing the approximation $\kappa_{d+1}=\Gamma[\frac{d-1}  {2}]/(2^{d+3}\pi^{(d-1)/2})$ \cite{Casini:2005zv} in large dimensions, we have
	\begin{equation}
		\label{eq.non}
		\frac{r_{o}}{\ell _{p}^{2}T}>\frac{d^{2}}{8y^{2}}\gg \frac{d^2}{8}   ,
	\end{equation}
	where $T$ is the Hawking temperature. Eq. (\ref{eq.non}) is a more general constraint that provides the limitation on the black hole size $r_{o}$ and temperature $T$, but note that the constraint (\ref{eq.non}) holds when $y \ll 1$. Moreover, if one sets charge $Q=0$ and $k=1$, the temperature changes to
	\begin{equation}
		T=\frac{d-2}{4 \pi r_{o}}.
	\end{equation}
	We can find that the constraint (\ref{eq.non}) can be written as $r_{o}/\ell_{p} \gg d^{3/2} /\sqrt{32 \pi e}$, which is the constraint of Schwarzschild black hole in large $d$ limit \cite{Du:2022vvg}.
	
	While in the extremal case, the constraint (\ref{eq5.8}) should be satisfied, equivalently
	\begin{equation}
		\label{eq5.17}
		\frac{d^{d}}{e^{d}}> 4\kappa _{d+1}cG_{N}\frac{(d-2)^{d}k^{\frac{d-1}{2}}}{r_{o}^{d-1} y^2 } \equiv \chi \kappa_{d+1}\ell_{p}^{d-1} \frac{(d-2)^{d}k^{\frac{d-1}{2}}}{r_{o}^{d-1}} ,
	\end{equation}
	where $\chi\equiv 4c/y^2$. In the large $d$ limit, we have
	\begin{equation}
		\chi \kappa_{d+1}\ell_{p}^{d-1} \frac{(d-2)^{d}k^{\frac{d-1}{2}}}{r_{o}^{d-1}} \simeq \chi \frac{d^{\frac{3d}{2}}\ell_{p}^{d}k^{\frac{d}{2}}}{8^{\frac{d}{2}}\pi^{\frac{d}{2}}e^{\frac{d}{2}}r_{o}^{d}},
	\end{equation}
	then by reorganizing eq. (\ref{eq5.17}), we obtain
	\begin{equation}\label{eq6.4.1}
		\frac{r_{o}^{2}}{\ell _{p}^{2}}> \frac{e k}{8\pi } \chi^{\frac{2}{d}}d \simeq  \frac{e}{8\pi }\left (1+\frac{r_{o}^{2}}{L^{2}}\right) d ,
	\end{equation}
	where we have used $\chi^{\frac{2}{d}}\to 1$ when $d \to \infty $ ( as we take a finite $y$ even it is small). To find a constraint on $r_{o}$, we write the above inequality as
	\begin{equation}
		\left(\frac{8\pi }{e \ell_{p}^{2}d }-\frac{1}{L^{2}} \right)r_{o}^{2}-1>0,
	\end{equation}
	which reproduces a quadratic inequality of variable $r_{o}$. Note that this inequality further requires
	\begin{equation}
		\frac{8\pi }{e \ell_{p}^{2} d } > \frac{1}{L^{2}},
	\end{equation}
	which puts a constraint on value of the AdS radius $L$, and $r_{o}$ should satisfy the follow relation
	\begin{equation}
		\label{eq5.2.2}
		\frac{r_{o}}{\ell _{p}}>\frac{1}{\sqrt{\frac{8\pi }{e d }-\frac{\ell_{p}^{2}}{L^{2}}}} >\sqrt{\frac{e}{8\pi}}d^{\frac{1}{2}}.
	\end{equation}
	This indicates that there is a universal lower bound on the radius of the large $d$ extremal RN-AdS$_{d+1}$ black hole. \omits {For the extremal large $d$ RN black hole ($k = 1$), from eq. (\ref{eq6.4.1}) we have $r_{o}/\ell_{p} > \sqrt{8 \pi/e }\cdot d^{1/2}$.}
	
	In short, we find new constraints on the large $d$ RN-AdS$_{d+1}$ black hole in the presence of island. Similar results have been found in Schwarzschild black hole in Refs. \cite{Du:2022vvg,Holdt-Sorensen:2019tne}. In Ref. \cite{Holdt-Sorensen:2019tne}, the authors provided a constraint, i.e. $r_{o}/\ell_{p} \gtrsim d^{3/2}$, on the size of Schwarzschild black hole through the large $d$ analysis. Instead, in Ref. \cite{Du:2022vvg} the authors also provided a constraint by the existence of the island, i.e. $r_{o}/\ell_{p}\gg d^{3/2}/\sqrt{32\pi e}$ for the large dimensional Schwarzschild black hole. In the present paper, we focused on finding new constraints in large $d$ limit RN-AdS$_{d+1}$ black hole. Indeed, we find constraints (\ref{constraint}) for the non-extremal case and  constraint (\ref{eq5.8}) for the extremal case, which leads to the constraints on the size of the large $d$ RN-AdS$_{d+1}$ black hole, i.e. eq. (\ref{eq.non}) and eq. (\ref{eq5.2.2}) severally. It is interesting to note that the constraint on $r_{o}$ for the extremal case is scaling as $d^{1/2}$, which is very different from that in Schwarzschild black hole \cite{Du:2022vvg,Holdt-Sorensen:2019tne}.

	\begin{figure}[htbp]
		\centering
		\includegraphics[width=0.8\columnwidth,height=0.5\linewidth]{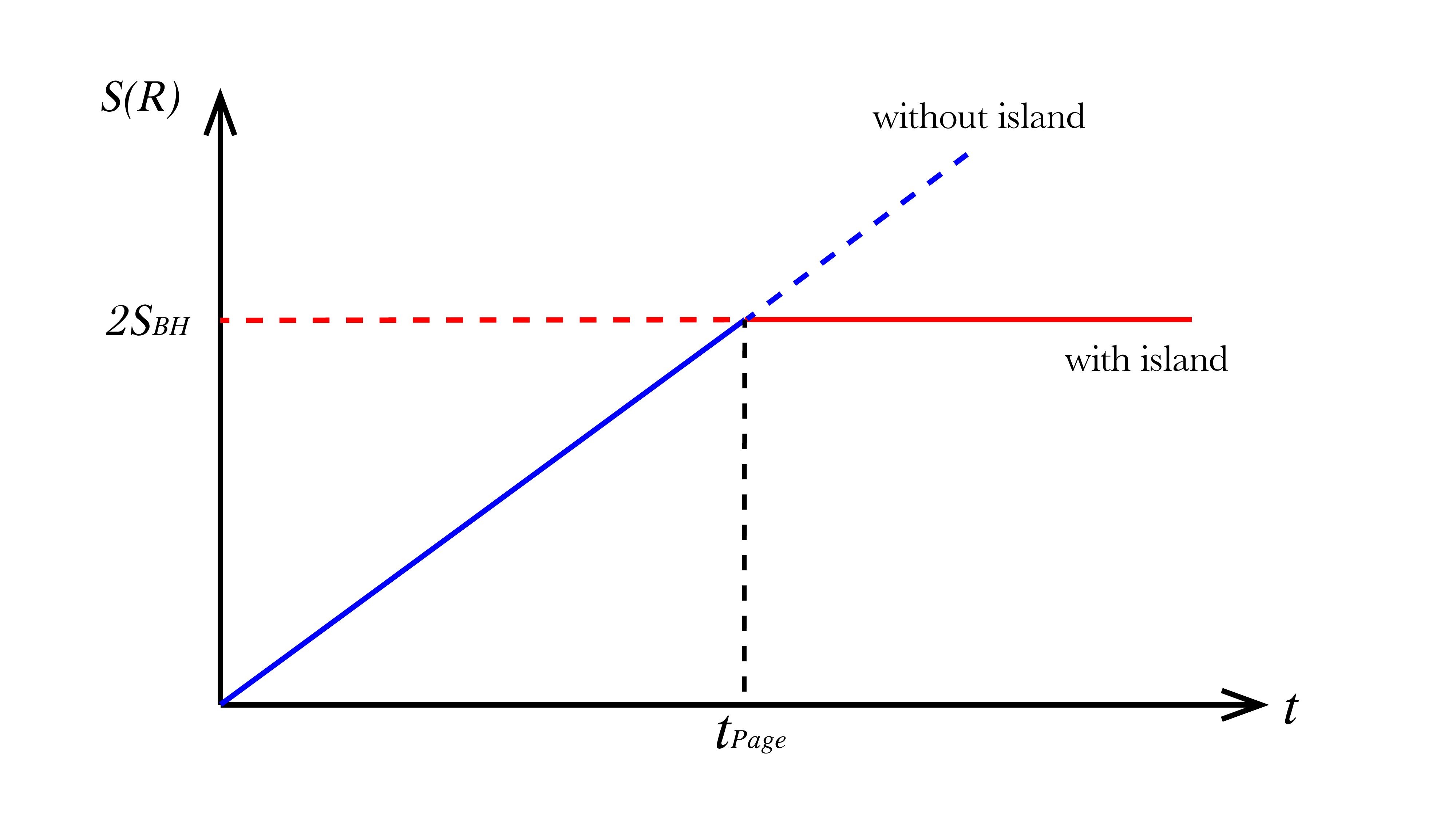}
		\caption{The Page curve for the non-extremal large dimensional eternal RN-AdS$_{d+1}$ black hole. For the eternal black hole, the entanglement entropy of Hawking radiation remains unchanged and constrained by twice Bekenstein-Hawking entropy after the Page time.}
		\label{fig.8}
	\end{figure}

	\section{Page curve and Page time}
	\label{sec.7}
	
	In this section, we would like to estimate the Page time. For the non-extremal case with $x < y \ll 1$, the entanglement entropy $S_{R}$ without island is given by eq. (\ref{eq4.3}), which grows linearly with $t$ at late time. While the entanglement entropy $S_{R}$ with island is given by eq. (\ref{eq4.2.1}), approximatively as $S_{R}\simeq 2S_{BH}$. This helps us reproduce the Page curve of the non-extremal large dimensional eternal RN-AdS$_{d+1}$ black hole (see the Fig. \ref{fig.8}). So the Page time in this case is given by
	\begin{equation}
		\label{eq6.1}
		t_{Page}\simeq\frac{6 S_{BH}}{c \kappa_{+}}=\frac{3S_{BH}}{\pi c T} ,
	\end{equation}
	where $T$ is the Hawking temperature of the non-extremal large $d$ RN-AdS$_{d+1}$  black hole.
		
For the extremal case with $x < y \ll 1$, although we have analyzed the Penrose diagram with island and give its entanglement entropy $S_{R}\simeq S_{BH}$, the entanglement entropy $S_{R}$ without island is ill-defined.
	This is mainly because $f(\rho)$ is ill-defined at the singularity $b_{0}=(t_{b}, 0)$. Therefore, we can not provide the Page time for the extremal black hole.\par
	
	In summary, for the non-extremal large $d$ RN-AdS$_{d+1}$ black hole, by combining eqs. (\ref{eq4.3}) and (\ref{eq4.25}), we get the Page curve of the entanglement entropy of Hawking radiation. Before the Page time, the entanglement entropy increases approximately linearly with time, and there is no island. After the Page time, the island appears and its boundary is near the horizon,  thus the entanglement entropy becomes approximately twice of the Bekenstein-Hawking entropy. The Page time is obtained in non-extremal case, which is the same as
	\begin{table}[htbp]
	\renewcommand{\arraystretch}{1.7}
	\centering
	\caption{The summary of results for large dimensional RN-AdS$_{d+1}$ black hole. }
	\begin{tabular}{cp{4cm}<{\centering}cp{3cm}<{\centering}cp{3cm}<{\centering}cp{3cm}<{\centering}}
		\toprule
		Black holes             & Without Island & With Island & Page time   \\ \midrule
		Non-extremal case & $S_{R}\simeq \frac{c}{3}\kappa_{+} t_{b}$           & $S_{R}\simeq 2S_{BH}$     & $t_{Page}\simeq \frac{3S_{BH}}{\pi c T}$         \\
		Extremal case     & ill-defined    & $S_{R}\simeq S_{BH}$      & ill-defined \\ \bottomrule
	\end{tabular}
	\label{tab.1}
	\end{table}
	 the result in Reissner-Nordstr{\"o}m black hole in four dimensions~\cite{Wang:2021woy}. While for the extremal case, we can not give a well-defined Page time (See the Tab. \ref{tab.1}).
	\
\section{Conclusion and Discussion}
\label{sec.8}
In this paper, we investigated the entanglement entropy of the Hawking radiation in the non-extremal and extremal cases of large $d$ RN-AdS$_{d+1}$ black hole coupled an auxiliary bath at the boundary of black hole via the island formula. We mainly considered the situation in which the boundary of the radiation region is close to the outer horizon of the black hole.
	
For the non-extremal case, we showed the existence of the general island solution, and we obtained two analytical island solutions in special cases $x \ll y/d $ and $x\sim y/d $ with $d \to \infty$, i.e. eq. (\ref{eq4.18}) and eq. (\ref{eq4.21}). Although we did not give the analytical expression of the island solution for general $x < y \ll 1$, we found a constraint (\ref{constraint}) that is required by the existence of the island in this case. The entanglement entropy of Hawking radiation has been obtained for both cases with and without the island. Meanwhile, the Page curve and Page time are also obtained (see the Fig. \ref{fig.8}).
	
For the extremal case, we showed that the entanglement entropy without island is ill-defined. As shown in the Penrose diagram (i.e. Fig. \ref{fig.6.1}), where the region extends from the boundary of the radiation region to the singularity $\rho =0 $, while the conformal factor (see eq. (\ref{eq5.1})) is divergent at $\rho =0 $. It has been pointed out in Ref.~\cite{Carroll:2009maa} that when taking the extremal limit for non-extremal RN black holes, the black hole geometry will divides into an extremal black hole and a disconnected AdS$_2$ part. While the microscopic entropy of the extremal RN black hole was shown can be calculated from its near horizon geometry either from the RN/CFT correspondence~\cite{Chen:2009ht}, or from the HEE perspective~\cite{Azeyanagi:2007bj}. However, the island formula will involve the black hole singularity in the absence of island for the extremal case, which will cause the semi-classical calculation to be invalid when the left-boundary reaches the singularity $b_{0}$ (see Fig. \ref{fig.6.1}). In previous works such as \cite{Kim:2021gzd,Ahn:2021chg,HosseiniMansoori:2022hok}, the authors mainly studied the case where the boundary radiation region is far from the outer horizon and utilized the formula (\ref{eq3.1}) for island phase. This differs from our analysis, as we mainly used the formula (\ref{eq3.2}) to calculate the entanglement entropy with island phase. We showed that the entanglement entropy of Hawking radiation in the extremal case is approximately equal to the Bekenstein-Hawking entropy ($S_{R} \simeq S_{BH}$), which is consistent with the results in previous studies \cite{Kim:2021gzd,Ahn:2021chg,HosseiniMansoori:2022hok}. However, since the entanglement entropy without island is not yet clear in the extremal case, therefore, the Page curve and Page time are also not well defined in this case.
	
Moreover, we showed that the existence of the island will put some constraints on the black holes (also see related papers \cite{Arefeva:2021kfx,Matsuo:2020ypv,Du:2022vvg,He:2021mst}). For large $d$ RN-AdS$_{d+1}$ black hole, we found the constraints both for the non-extremal and extremal cases, which are
	\begin{equation}
		\label{eq.cnonext}
		\frac{r_{o}}{\ell _{p}^{2}T}>\frac{d^{2}}{8y^{2}}\gg \frac{d^2}{8}   \quad  (\rm{for\quad nonextremal\quad case}),
	\end{equation}
	\begin{equation}
		\label{eq.cext}
		\frac{r_{o}}{\ell _{p}}>\frac{1}{\sqrt{\frac{8\pi }{e d }-\frac{\ell _{p}^2}{L^{2}}}} >\sqrt{\frac{e}{8\pi}}d^{\frac{1}{2}} \quad \text{and}  \quad \frac{L}{\ell_{p}} > \sqrt{\frac{e}{8\pi}}d^{\frac{1}{2}}\quad (\rm{for\quad extremal\quad case}),
	\end{equation}
	as required by the existence of island in the case $y\ll 1$.
	
Now let us discuss some issues needed to be solved in the future. Firstly, we focused on studying the case of one island in our paper. In general, the configuration of multiple islands is allowed, and it can soften the turning point of the Page curve at the Page time. Secondly, our calculation was mainly based on the two dimensional approximation formula in Sec. \ref{sec.3}, a more general formula is needed to calculate the entanglement entropy of matter in the high dimensional curved spacetime. Finally, the constraints we obtained are only valid in the case of $x<y\ll 1$. For more general case $y> 0$, we do not yet know whether there is a similar constraint, it deserves to generalize the calculation of the entanglement entropy in high-dimensional black holes for general case $y> 0$.




\acknowledgments
	
This work was supported by the National Natural Science Foundation of China (No.~11675272).

	
	
	
	
	\bibliographystyle{JHEP}
	\bibliography{ref}
\end{document}